\documentclass[11 pt]{article}
\pdfoutput=1 
\usepackage{mathrsfs}
\usepackage{jheppub} 
\usepackage{mathtools,amsmath,amssymb}
\allowdisplaybreaks      
\newcommand{\ep}{\epsilon}
\newcommand{\sig}{\sigma}
\newcommand{\mpow}[1]{m_p^{#1}}
\newcommand{\SQ}{\sqrt{\sig^2-1}}

\usepackage{amssymb}
\usepackage{dsfont}
\usepackage{amsmath}
\usepackage{bbm}
\usepackage{amsfonts}
\usepackage{pifont}
\usepackage{orcidlink}
\usepackage{multicol} \usepackage[version=4]{mhchem} \usepackage{multirow}
\usepackage[customcolors,shade]{hf-tikz}
\usepackage{graphicx}
\usepackage{tikz}
\usepackage{tikz-feynman}
\tikzfeynmanset{compat=1.1.0}
\usepackage{physics}
\usepackage[most]{tcolorbox}
\usepackage{amsmath}
\usepackage{caption}
\usepackage{subcaption,longtable,stmaryrd}
\usepackage{slashed}
\usepackage{bigints}
\usepackage{cancel}
\usepackage{multicol}
\usepackage{blindtext}
\usepackage{tikz}
\usepackage{enumitem}
\usepackage[customcolors,shade]{hf-tikz}
\newcommand{\penguinsymbol}[1]{%
  \begin{tikzpicture}[x=1ex,y=1ex,scale=#1,baseline=-0.2ex]
    \fill[black]
      (-0.7,-1.0) .. controls (-0.9,-0.2) and (-0.6,0.9) .. (0,1.1)
      .. controls (0.6,0.9) and (0.9,-0.2) .. (0.7,-1.0) -- cycle;
    \fill[white] (0,-0.2) ellipse (0.45 and 0.7);
    \fill[black] (-0.9,-0.2) .. controls (-1.4,0.2) and (-1.1,-0.6) .. (-0.7,-0.9) -- cycle;
    \fill[black] ( 0.9,-0.2) .. controls ( 1.4,0.2) and ( 1.1,-0.6) .. ( 0.7,-0.9) -- cycle;
    \fill[white] (-0.18,0.75) circle (0.12);
    \fill[white] ( 0.18,0.75) circle (0.12);
    \fill[black] (-0.18,0.75) circle (0.05);
    \fill[black] ( 0.18,0.75) circle (0.05);
    \fill[orange] (0,0.55) -- (-0.12,0.40) -- (0.12,0.40) -- cycle;
    \fill[orange] (-0.30,-1.05) -- (-0.60,-1.25) -- (-0.10,-1.25) -- cycle;
    \fill[orange] ( 0.30,-1.05) -- ( 0.60,-1.25) -- ( 0.10,-1.25) -- cycle;
  \end{tikzpicture}%
}

\DeclareRobustCommand{\penguin}{%
  \mathchoice{\penguinsymbol{1.10}}
             {\penguinsymbol{1.00}}
             {\penguinsymbol{0.80}}
             {\penguinsymbol{0.65}}
}
\usepackage{graphicx}
\usepackage{cancel}
\DeclareSymbolFont{usualmathcal}{OMS}{cmsy}{m}{n}
\DeclareMathAlphabet\mathbfcal{OMS}{cmsy}{b}{n}
\DeclareSymbolFontAlphabet{\mathcal}{usualmathcal}

\DeclareSymbolFont{rmlargesymbols}{OMX}{mdbch}{m}{n}
\DeclareMathSymbol{\intop}{\mathop}{rmlargesymbols}{82}
\DeclareMathSymbol{\rmointop}{\mathop}{rmlargesymbols}{72}
\newcommand{\filledsquare}[1]{\tikz[baseline=-0.5ex]\draw[fill=#1, draw=black] (0,0) rectangle (0.8ex,0.8ex);}

\definecolor{mygray}{gray}{0.5}

{\title{\boldmath{Heterotic Footprints in Classical Gravity: PM dynamics from On-Shell soft amplitudes at one loop }}}

\author[]{Arpan Bhattacharyya,}
\author[]{Saptaswa Ghosh,}
\author[]{Ankit Mishra,} 
\author[]{Sounak Pal}

\affiliation[]{Indian Institute of Technology Gandhinagar, Palaj, Gujarat: 382355}

\emailAdd{abhattacharyya@iitgn.ac.in}
\emailAdd{saptaswaghosh@iitgn.ac.in}
\emailAdd{ankit.mishra@iitgn.ac.in}
\emailAdd{palsounak@iitgn.ac.in}

\abstract{We study classical scattering of charged black holes in Einstein-Maxwell-Dilaton (EMD) theory. Working in the classical (Post-Minkowskian) regime, we extract the conservative two-body potential by expanding the one-loop amplitudes in the soft regime. We explicitly show that, as in GR, the relevant soft amplitudes are infrared (IR) finite when long-range interactions are consistently treated via the Lippmann-Schwinger equation and the associated IR subtraction. The scattering angle is then obtained from the eikonal exponentiation of the soft amplitude. Our results track the separate roles of electromagnetic and dilatonic charges in both the conservative dynamics and the eikonal phase, and they smoothly reduce to the GR limit when the charges and the dilaton coupling are switched off. Where applicable, we compare our results with existing literature and find agreement. These findings provide amplitude-based benchmarks for compact-object dynamics in EMD and furnish building blocks for waveform modeling in beyond-GR scenarios.}
\makeatletter
\begin{document}
\maketitle
\section{Introduction}
Networks of ground-based laser interferometers have inaugurated gravitational-wave astronomy, offering a new way to observe to the Universe in the high-frequency band \cite{LIGOScientific:2016aoc,LIGOScientific:2016sjg,LIGOScientific:2016vlm,LIGOScientific:2017bnn,LIGOScientific:2019hgc}. Forthcoming space-based missions will open the low-frequency window, and pulsar-timing arrays are poised to probe gravitational waves in the nanohertz regime. Beyond their astrophysical returns, these observations enable stringent tests of Einstein’s general relativity. Whereas many early studies placed theory-agnostic bounds on deviations from GR, current efforts increasingly construct waveform templates within specific modified-gravity frameworks. The LIGO/Virgo observations of black-hole and neutron-star inspirals and mergers, in particular, require high-precision analytic calculations of both the conservative dynamics (classical potential, scattering angle, spin-kick) and the emitted radiation from compact binaries \cite{Purrer:2019jcp}. Methodologically, this level of precision parallels scattering cross-section computations in particle physics and naturally motivates the use of quantum-field-theoretic tools. In this spirit, perturbative quantum-gravity techniques framed in QFT language have proven highly effective for studying the classical gravitational interactions of black holes and neutron stars.\\\\
General relativity is the most successful classical theory of gravitation, explaining how massive bodies curve spacetime and how this curvature governs the motion of matter and light. Nonetheless, GR is widely regarded as incomplete at very small length scales (high energies)-the “UV” regime—where quantum effects become essential, i.e., GR is not a UV-complete theory of gravity. This motivates the search for a consistent, UV-complete quantum theory of gravity. Within a low-energy, model-independent framework, Weinberg advocated treating gravity as an effective field theory, augmenting the Einstein–Hilbert action by an infinite tower of higher-derivative operators \cite{PhysRev.138.B988}. These additions can be viewed as higher-curvature modifications of GR. A complementary route is to enlarge the field content by introducing additional degrees of freedom such as scalar or vector (gauge) fields alongside the graviton, with the potential to address phenomena associated with dark matter and dark energy \cite{Damour:1992we,Horbatsch:2015bua,Schon:2021pcv,Rainer:1996gw,DeFelice:2011bh, Alviani:2025xvf}. In principle, higher–curvature corrections and additional degrees of freedom (DOFs) can coexist, producing broad classes of modified–gravity theories. Although extra DOFs are often introduced phenomenologically, string theory supplies them from first principles: in the low–energy limit (e.g., of the heterotic string) the massless sector necessarily contains the graviton \(g_{\mu\nu}\), dilaton \(\theta\), antisymmetric tensor \(B_{\mu\nu}\), and non-Abelian gauge fields \(A_\mu^{I}\). For backgrounds that are independent of \textcolor{black}{\(d\) of the spacetime directions} and for gauge configurations commuting with \(p\) Abelian generators, the space of classical solutions enjoys a continuous \(O(d)\times O(d+p)\) symmetry that mixes \(g_{\mu\nu}\), \(B_{\mu\nu}\), \(\theta\), and \(A_\mu^{I}\) and acts as a solution-generating transformation \cite{Hassan:1991mq}. Acting with this “twist,” one can, for example, start from a ten-dimensional black 6-brane carrying magnetic charge but no electric or antisymmetric-tensor gauge charge and generate a family of inequivalent solutions with independent electric, magnetic, and antisymmetric-tensor charges \cite{Hassan:1991mq}. Also in this framework, additional DOFs and higher–derivative corrections arise in tandem rather than as ad hoc additions, yielding symmetry–guided targets for EFT analyses and phenomenology. The ensuing low–energy effective field theory admits exact charged (and spinning) black–hole solutions \cite{GIBBONS1988741,Garfinkle:1990qj,Sen:1992ua}, motivating detailed studies of their classical dynamics as probes of potential stringy imprints in gravitational-wave observations.\\\\
Several classical frameworks for modeling black-hole and neutron-star binaries, covering inspiral and merger, have matured in recent years \cite{Blanchet:2013haa, Schafer:2018kuf,cite-key, Pati:2000vt}. Using traditional methods, the conservative gravitational potential and waveform emission have been worked out through 3.5PN order \cite{Blanchet:1995ez,Tagoshi:2000zg,Blanchet:2004ek,Faye:2006gx,Blanchet:2006gy,Damour:2000ni,Itoh:2003fy,Blanchet:2001ax}, with subsequent advances pushing the state of the art to 4.5PN \cite{Blanchet:2023sbv,Blanchet:2023bwj}. This program has been extended to incorporate finite-size effects—most notably spins and tidal deformability—thereby refining the binary dynamics \cite{PhysRevD.12.329,Kidder:1992fr,Cho:2022syn,Steinhoff:2007mb,Arun:2008kb,Henry:2020ski,Mandal:2024iug}. Parallel developments have carried these analyses beyond General Relativity, yielding extensive results in a variety of modified-gravity settings \cite{Zhang:2017srh,Ma:2019rei,Bernard:2022noq,AbhishekChowdhuri:2022ora,Higashino:2022izi,Zhang:2018prg,Li:2022grj,Shiralilou:2021mfl,Kumar:2023bdf,Trestini:2024mfs}.
\\\\
In tandem, quantum-field-theoretic (QFT) techniques have emerged as highly efficient tools for the classical two-body problem. For slowly moving inspirals, the post-Newtonian (PN) expansion remains well suited, and a broad EFT literature has systematized these calculations \cite{Goldberger:2004jt,Goldberger:2009qd,Kol:2007bc,Goldberger:2007hy,Porto:2016pyg,Foffa:2013qca,cite-key2,Levi:2018nxp}, with extensions beyond GR—including extra degrees of freedom and higher-curvature EFTs—explored in \cite{Bhattacharyya:2023kbh, Bhattacharyya:2025slf}. By contrast, for hyperbolic encounters and high-velocity events \cite{hyp,hyp1,hyp2,hyp3,hyp4,hyp7,hyp8,hyp9,hyp13,hyp16,hyp14,Usseglio:2025iwt}, the PN approximation is inadequate, and the Post-Minkowskian (PM) framework becomes the natural language. In GR, the PM literature is extensive \cite{Damour:2016gwp,Bini:2017wfr,Bini:2017xzy,Damour:2017zjx,Damour:2019lcq,Bini:2020flp,Bini:2020uiq,Damour:2020tta,Bini:2020rzn,Bini:2021gat,Bini:2022enm,Damour:2022ybd,Bini:2022wrq,Rettegno:2023ghr,Bini:2023fiz}, and several complementary field-theory approaches have crystallized: direct scattering-amplitude methods \cite{Brandhuber:2022qbk,Brandhuber:2023hhy,Kosower:2018adc,Bjerrum-Bohr:2018xdl,Bjerrum-Bohr:2013bxa,Alessio:2024wmz, Bern:2019crd, Bern:2019nnu, Bern:2020buy,Bern:2024adl,Bern:2023ity,Bern:2022kto, Bern:2021yeh,Bern:2021dqo,Bern:2020gjj,Bern:2025zno, Brandhuber:2019qpg,AccettulliHuber:2019jqo,AccettulliHuber:2020oou,Bjerrum-Bohr:2021wwt,Bjerrum-Bohr:2021vuf,Bjerrum-Bohr:2020syg, Bjerrum-Bohr:2019kec, Cristofoli:2019neg,Bjerrum-Bohr:2016hpa,Chen:2024mmm,Alessio:2022kwv,Alessio:2023kgf, ashoke1, Gonzo:2024zxo,Aoude:2023vdk,Aoude:2023dui,Adamo:2024oxy,Adamo:2021rfq,Georgoudis:2023eke,Georgoudis:2024pdz,Bini:2024rsy, DeAngelis:2023lvf, Cristofoli:2021vyo,Mougiakakos:2022sic}, the worldline EFT formulation \cite{Kalin:2020mvi, Kalin:2020lmz,Dlapa:2024cje,Dlapa:2023hsl,Dlapa:2025biy}, and, most recently, worldline quantum field theory (WQFT) \cite{Mogull:2020sak,Jakobsen:2021lvp,Jakobsen:2023ndj,Jakobsen:2023hig,Jakobsen:2022psy,Wang:2022ntx, Driesse:2024feo}. Extensions of these PM and field-theoretic methods to scenarios beyond GR have also begun to appear \cite{Cristofoli:2019ewu,Bhattacharyya:2024aeq, Bhattacharyya:2024kxj,Brandhuber:2024bnz,Falkowski:2024bgb,Falkowski:2024yuy,Wilson-Gerow:2025xhr}.
\\\\
In this work, we adopt the scattering–amplitude approach to compute the conservative two–body potential, the eikonal phase, and, consequently, the scattering angle through one loop. Following \cite{Parra-Martinez:2020dzs}, we organize the calculation in terms of soft amplitudes and extract the angle via momentum–space eikonal exponentiation. In parallel, we derive the two–body potential from the Lippmann–Schwinger equation and show that, after performing the appropriate EFT (Born) subtraction of long–range iterations, the momentum–space potential is infrared (IR) finite. The paper is organized as follows. In section~\eqref{sec2}, we motivate Einstein--Maxwell--Dilaton (EMD) theory as the low–energy effective description of the heterotic string, fix conventions and gauge choices, and derive the propagators and minimal set of on–shell Feynman rules required for $2\!\to\!2$ scattering. In section~\eqref{sec3}, we develop the analytic toolkit for the soft regime: using dimensional regularization, expansion by regions, and integration–by–parts (IBP) reduction, we obtain a compact basis of master integrals and their soft–limit expansions that capture the classical contributions. With these ingredients, we assemble the complete one–loop amplitudes across the relevant topologies (single exchanges, triangles, boxes, penguins) in the classical limit. In section~\eqref{sec4}, we extract the conservative two–body potential via the Lippmann-Schwinger equation and demonstrate that, after a careful EFT/Born subtraction of long–range iterations, the momentum–space potential is infrared finite. Section~\eqref{sec5} then determines the eikonal phase by exponentiating the momentum–space amplitude and, consequently, the scattering angle through one loop; we verify the smooth GR limit and delineate the separate roles of electromagnetic and dilatonic charges. We conclude by comparing with existing literature and commenting on the implications for amplitude–based dynamics beyond GR.

\section{Setup and methodology}\label{sec2}
\textbf{Low energy effective action of Heterotic string and emergence of EMD theory:}\\\\
The low-energy effective action of Heterotic string theory in string frame is given by \cite{Metsaev:1987zx,Hassan:1991mq},
\begin{align}
    \begin{split}
        S_{\textrm{low-eff}}=\int d^4x \sqrt{-G}\,e^{-\theta}\left[-R+\frac{1}{12}H_{\mu\nu\rho}H^{\mu\nu \rho}+G^{\mu\nu}\partial_{\mu}\theta \partial_{\nu}\theta-\frac{1}{8}F^2\right]
    \end{split}
\end{align}
    where $G_{\mu\nu}$ is the metric, $F_{\mu\nu}=\partial_\mu A_{\nu}-\partial_{\nu}A_{\mu}$ is the field strength corresponding to the Maxwell field $A_{\mu}$ and $\theta$ is the dilaton field and , $H_{\mu\nu\rho}=\partial_{\mu} B_{\nu\rho}+\textrm{cyclic per.}-(\Omega_3(A))_{\mu\nu\rho}$, where, $B_{\mu\nu}$ is the antisymmetric tensor field and $(\Omega_3(A))_{\mu\nu\rho}=\frac{1}{4}(A_{\mu}F_{\nu\rho}+\textrm{cyclic per.})$ is the gauge Chern-Simons term. The theory has several features to note. First, it arises from heterotic string theory, compactified from ten to four dimensions. In the compactification procedure, massless fields that were included in the effective action arise. Also, only $U(1)$ subsector of full $E_{8}\times E_{8}$ or $\texttt{Spin}(32)/\mathbb{Z}_2$ has been included. The metric $G_{\mu\nu}$ (naturally appears in the $\sigma-$ model) is conformally related to the Einstein metric $g_{\mu\nu}$. Lastly, the action has been truncated upto the second order derivative of the field content. Now, ignoring the antisymmetric field $H_{\mu\nu\rho}$ and using proper conformal field redefinition, the effective action takes the following form (restoring the factors of $m_p$),
    \begin{align}
        \begin{split}
              S_{\textrm{low-eff}}=\int d^4 x \sqrt{-g}\left[-\frac{m_p^2}{2}R+\frac{1}{2}g^{\mu\nu}\partial_{\mu}\theta\,\partial_\nu \theta-\frac{1}{16}\,e^{-\sqrt{2}\frac{\theta}{m_p}}F^2\right]\,.
        \end{split}
    \end{align}
The dilaton field $\theta$ plays a central role in the theory as its vacuum expectation value is related to the coupling of the theory as $\texttt{g}_{c}=e^{\sqrt{2}\langle\theta\rangle}$. The charged black hole solution (in unit of $G_N=1$) takes the following form \cite{Garfinkle:1990qj,GIBBONS1988741},
\begin{align}
\begin{split}
   & ds^2=\left(1-\frac{2M}{r}\right)dt^2-\left(1-\frac{2M}{r}\right)^{-1}dr^2-r\left(r-\frac{Q^2e^{-\sqrt{2}\theta_0}}{M}\right)d\Omega\,,
  \\ & e^{-\sqrt{2}\theta}=e^{-\sqrt{2}\theta_0}\left(1-\frac{Q^2\, e^{-\sqrt{2}\theta_0}}{Mr}\right)\,.
  \end{split}
\end{align}
Now, expanding the action around $\theta=\langle \theta\rangle:=\theta_0$ we  get,
     \begin{align}
        \begin{split}
              S_{\textrm{low-eff}}&=\int d^4 x \sqrt{-g}\left[-\frac{m_p^2}{2}R+\frac{1}{2}g^{\mu\nu}\partial_{\mu}\delta\theta\,\partial_\nu \delta\theta-\frac{1}{16}e^{-\sqrt{2}\theta_0/m_p}\,e^{-\sqrt{2}\frac{\delta\theta}{m_p}}F^2\right]\,, \\ &
              =\int d^4 x \sqrt{-g}\left[-\frac{m_p^2}{2}R+\frac{1}{2}g^{\mu\nu}\partial_{\mu}\delta\theta\,\partial_\nu \delta\theta-\frac{1}{16 \,\texttt{g}_{c}}\,e^{-\sqrt{2}\frac{\delta\theta}{m_p}}F^2\right]\,. \label{1.3h}
        \end{split}
    \end{align}
The theory described in \eqref{1.3h} has exact charged blackhole solution. The main goal of the paper is to study blackhole scattering and investigate how string theory modifies the fundamental classical observables such as the scattering angle (and the classical potential). As a standard procedure, we model the black holes by a charged scalar field $\Phi$ with dilaton dependent effective mass ($m_i(\theta)$)  , and the matter action is given by,
\begin{align}
    \begin{split}
        S_{\textrm{matter}}=\int d^4x \sqrt{-g}\sum_{i=1}^2\left[g^{\mu\nu}D_{\mu}\Phi^{\dagger}_iD_{\nu}\Phi_i-m_i^2(\theta)\Phi^{\dagger}_i\Phi_i\right]
    \end{split}
\end{align}
where the covariant derivative $D_{\mu}=\partial_{\mu}-ie\alpha_iA_{\mu}$ encodes the information about the gauge field $A_{\mu}$ and the electric charge of the scalar field $\Phi$, \textcolor{black}{ $Q_i=\alpha_i\,e$}.
Note that in the presence of the dilaton, the mass of the scalar field is now a function of the dilaton $\theta$. For computational purposes, the mass can be expanded around the background/asymptotic value of dilaton, i.e. $\theta_0$ as,
\begin{align}
    m_i(\theta)=m_i(\theta_0)+a_i(\theta_0)\,\delta\theta+b_i (\theta_0)\delta\theta^2+\cdots.
\end{align}
For notational simplicity, we use $\delta\theta\to \theta$, for further computation and define the path integral as, $$\int D\delta\theta\, D[A,g,\Phi_i]\cdots\to \int D\theta \,D[A,g,\Phi_i]\cdots\,. $$\\
\textbf{Propagators and Feynman rules:}\\
The two-point function for the Maxwell field will be modified due to the presence of string coupling $\texttt{g}_c$, but the two-point function of the dilaton and graviton remains unchanged. The gauge fixed action for the Maxwell field is given by,
\begin{align}
    \begin{split}
S_{\texttt{photon}}=\frac{1}{4\texttt{g}_c}\int d^4x \left(-\frac{1}{4}F^2-\frac{1}{2\xi}(\partial\cdot A)^2\right)\,.
    \end{split}
\end{align}
Immediately, we can identify the momentum space propagator in Feynman gauge as,
\begin{align}
\begin{split}
 \hspace{-7 cm}  D_{\mu\nu}(k)=\begin{minipage}[h]{0.12\linewidth}
	\vspace{4pt}
	\scalebox{1.4}{\includegraphics[width=\linewidth]{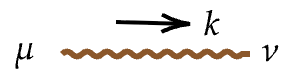}}
\end{minipage}\hspace{1.2 cm}=
  4\texttt{g}_c\frac{-i\,\eta_{\mu\nu}}{k^2+i0}.
    \end{split}
\end{align}
Similarly the graviton\footnote{We define $P_{\mu\nu;\rho\sigma}=\left(\frac{1}{2}\eta_{\mu\rho}\eta_{\nu\sigma}+\frac{1}{2}\eta_{\mu\sigma}\eta_{\nu\rho}-\frac{1}{d-2}\eta_{\mu\nu}\eta_{\rho\sigma}\right)$.} and dilaton propagator takes the form,

\begin{align}
\begin{split}
  & 
 D_{\mu\nu;\rho \sigma}(k)= \begin{minipage}[h]{0.12\linewidth}
	\vspace{0 pt}
	\scalebox{1.8}{\includegraphics[width=\linewidth]{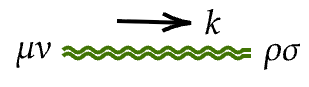}}
\end{minipage}\hspace{1.4cm}=\frac{i}{k^2+i0} \left(\frac{1}{2}\eta_{\mu\rho}\eta_{\nu\sigma}+\frac{1}{2}\eta_{\mu\sigma}\eta_{\nu\rho}-\frac{1}{d-2}\eta_{\mu\nu}\eta_{\rho\sigma}\right),\,\\ &
    D(k)=     \begin{minipage}[h]{0.12\linewidth}
	\vspace{4pt}
	\scalebox{1.4}{\includegraphics[width=\linewidth]{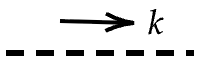}}
\end{minipage}\hspace{1.2 cm}=\frac{i}{k^2+i0}.
  \end{split}
\end{align}
\\
Now, expanding the bulk gravitational and matter actions about the background solution and truncating at cubic order yields the three-point interaction vertices needed for our analysis. We enumerate them below.\\\\
\textbf{\textit{Bulk-Matter interaction vertex:}}
\vspace{0 cm}
\begin{align}
    \begin{split}
      &\hspace{-0.2 cm}     \bullet\,\,\begin{minipage}[h]{0.12\linewidth}	
\scalebox{1.5}{\includegraphics[width=\linewidth]{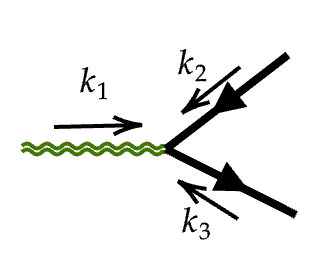}}
\end{minipage}\hspace{1 cm}
\equiv 
\hspace{0 cm}\,\frac{i}{m_p}\left(k_3^{(\mu}k_2^{\nu)}-\frac{1}{2}\left(k_2\cdot k_3+m_0^2\right)\eta^{\mu\nu}\right),
\bullet\,\, \begin{minipage}[h]{0.12\linewidth}	
\scalebox{1.5}{\includegraphics[width=\linewidth]{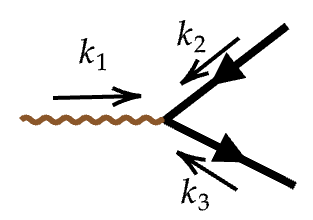}}
\end{minipage}\hspace{1 cm}
\equiv 
\hspace{0 cm}\, i\,e\, \alpha_i\,(k_3^\mu-k_2^\mu),\\ &
 \hspace{-0.2 cm}     \,\,\bullet\,\,\begin{minipage}[h]{0.12\linewidth}	
\scalebox{1.5}{\includegraphics[width=\linewidth]{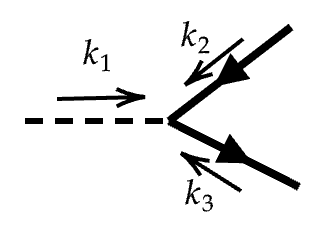}}
\end{minipage}\hspace{1 cm}
\equiv 
\hspace{0 cm}\, -i \frac{2m_0 a}{m_p},\,\bullet\, \begin{minipage}[h]{0.12\linewidth}	
\scalebox{1.5}{\includegraphics[width=\linewidth]{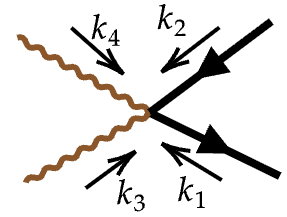}}
\end{minipage}\hspace{1 cm}
\equiv 
\hspace{0 cm}\,2 {ie}^2 \eta ^{\mu \nu },\,\,\bullet\,\hspace{-0.1 cm} \begin{minipage}[h]{0.12\linewidth}	
\scalebox{1.5}{\includegraphics[width=\linewidth]{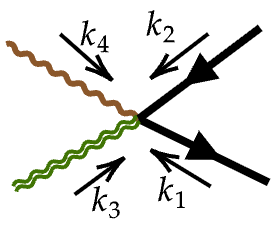}}
\end{minipage}\hspace{1 cm}
\equiv 
\hspace{0 cm}\,\frac{i e }{m_p}P^{\mu \nu ,\alpha \beta }(k_1-k_2)_{\beta },
    \end{split}
\end{align}
\begin{align}
    \begin{split}
    \\ &
\hspace{-0.2 cm}\bullet\, \begin{minipage}[h]{0.12\linewidth}	
\scalebox{1.5}{\includegraphics[width=\linewidth]{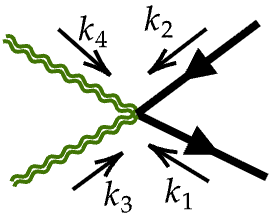}}
\end{minipage}\hspace{1 cm}
\equiv 
\hspace{0 cm}\,-\frac{i}{2 m_p^2}\Bigg[k_1^{\rho } k_2^{\sigma } \Big(-\eta _{\mu \nu } I_{\alpha \beta ,\rho \sigma }-\eta _{\alpha \beta } I_{\mu \nu ,\rho \sigma }+\eta _{\beta \nu } I_{\alpha \mu ,\rho \sigma }\\ &\hspace{5 cm}+\eta _{\alpha \mu } I_{\beta \nu ,\rho \sigma }+\eta _{\beta \mu } I_{\alpha \nu ,\rho \sigma }+\eta _{\alpha \nu } I_{\beta \mu ,\rho \sigma }\Big)-P_{\mu \nu ,\alpha \beta }(k_1 \cdot k_2+m^2)\Bigg],\\ &
        \hspace{-0.2 cm}\bullet \begin{minipage}[h]{0.12\linewidth}	
\scalebox{1.5}{\includegraphics[width=\linewidth]{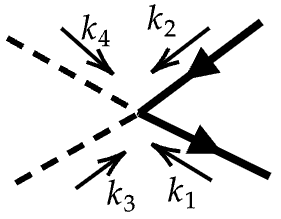}}
\end{minipage}\hspace{1 cm}
\equiv 
\hspace{0 cm}\,\frac{2 i \left(a^2-2 m_i\,b\right)}{m_p^2},\,\,\bullet \begin{minipage}[h]{0.12\linewidth}	
\scalebox{1.5}{\includegraphics[width=\linewidth]{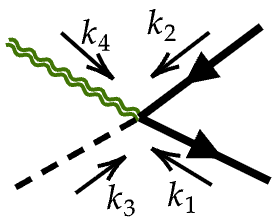}}
\end{minipage}\hspace{1 cm}
\equiv 
\hspace{0 cm}\,-\frac{i\,a  m_i \eta ^{\mu \nu }}{m_p^2}.
    \end{split}
\end{align}
\textbf{\textit{Bulk  interaction vertex:}}
\begin{align}
    \begin{split}
      &\hspace{0 cm}     \bullet\,\,\begin{minipage}[h]{0.12\linewidth}	
\scalebox{1.5}{\includegraphics[width=\linewidth]{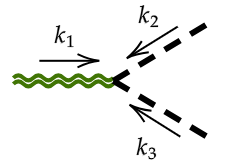}}
\end{minipage}\hspace{1 cm}
\equiv 
\hspace{0 cm}\,\frac{i}{4m_p}\left( k_1^\mu k_2^\nu + k_1^\nu k_2^\mu - k_1 \cdot k_2 ~\eta^{\mu \nu} \right)  ,
\\ &
\,\bullet\,\,\begin{minipage}[h]{0.12\linewidth}	
\scalebox{1.5}{\includegraphics[width=\linewidth]{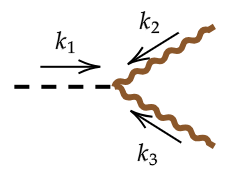}}
\end{minipage}\hspace{1 cm}
\equiv 
\hspace{0 cm}\, -i\frac{2\sqrt{2} }{16 \texttt{g}_c m_p}\left(\eta^{\mu\nu} k_2 \cdot k_3-k_2^{(\mu}k_3^{\nu)}\right),
\\ &
 \hspace{0 cm}     \bullet\,\,\begin{minipage}[h]{0.12\linewidth}	
\scalebox{1.5}{\includegraphics[width=\linewidth]{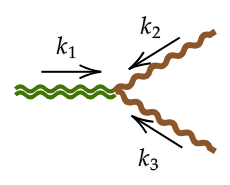}}
\end{minipage}\hspace{1 cm}
\equiv 
\hspace{0 cm}\,  -\frac{i  }{4\texttt{g}_c m_p}\Bigg[~  (k_2 \cdot k_3 ) P^{\alpha \beta , \mu \nu} +\eta^{\mu \nu} k_2^{ (\alpha} k_3^{\beta)}  
 + 
\frac{1}{2}\Big( \eta^{\alpha \beta}k_2^{\mu}k_3^{\nu} -\eta^{\alpha \mu}k_2^{\beta}k_3^{\nu} - \eta^{\alpha \nu}k_2^{\mu}k_3^{\beta}\\&\hspace{10 cm} -  \eta^{\beta \mu}k_2^{\alpha}k_3^{\nu} - \eta^{\beta \nu}k_2^{\mu}k_3^{\alpha} \Big)    \Bigg].     
    \end{split}
\end{align}
With all the ingredients in hand, we now go on to the computation of the scattering amplitude at the soft limit and discuss how this particular limit of the full scattering amplitude is enough to generate the classically relevant observables. However, we also show that at one loop it can also generate \textit{some part} of the quantum correction to the classical potential. 
\newpage
\section{Computation of scattering amplitude: Soft-loop expansion and classical limit}\label{sec3}
We first focus on the computation of the scattering amplitude in the soft limit \cite{Parra-Martinez:2020dzs} and then we go on to the computation of the Post-Minkowskian potential. A general $L$-loop, $n$-point amplitude takes the following form,
\begin{align}
\begin{split}
i \mathbfcal{A}_{n}^{L-\textrm{loop}}\left[\begin{minipage}
          [h]{0.15\linewidth}
	\vspace{1.2 pt}
	\scalebox{1}{\includegraphics[width=\linewidth]{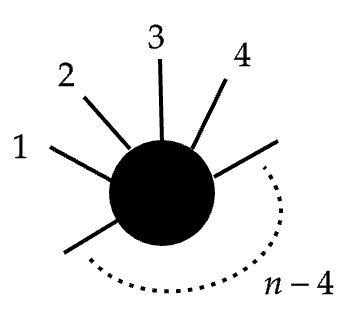}}
      \end{minipage}\right]=\sum_{j}\int \left(\prod_{l=1}^L\frac{d^d\ell_{l}}{(2\pi)^D}\right) \frac{1}{S_{j}}\frac{\mathcal{N}_{j}}{\prod_{\gamma_j}\rho_{\gamma_j}}
      \end{split}
\end{align}
where the sum runs over all possible L-loop Feynman diagrams $j$, for each diagram $\ell_{l}$ denotes the L loop momenta, $\gamma_j$ labels the propagators, and $S_j$ is associated with the symmetry factor of the diagram. $n_j's$ are the polynomials constructed from the irreducible scalar products constructed from external and loop momenta. In the subsequent computations, we will focus on the one-loop, four-point scattering amplitude, $\mathbfcal{A}_{4}^{1-\textrm{loop}}$, which is relevant for a binary scattering event. At the classical limit, any one-Loop, four-point amplitude can be written as,
\begin{align}
\begin{split}
   \mathbfcal{A}_{4}^{1-\textrm{loop}}={\mathfrak{c}_{\square}\mathcal{A}_{\square}+\mathfrak{c}_{\Join}\mathcal{A}_{\Join}+\mathfrak{c}_{\triangle}\mathcal{A}_{\triangle}+\mathfrak{c}_{\triangledown}\mathcal{A}_{\triangledown}+(\textrm{purely quantum})}\,.
   \end{split}
\end{align}
After integrating, it takes the form,
\begin{align}
     \mathbfcal{A}_{4}^{1-\textrm{loop}}=A+B \,q^2+\cdots+\frac{\alpha}{\sqrt{-q^2}}+\beta \log(\sqrt{-q^2})+\cdots\,.
\end{align}
Interestingly, it is trivial to show (by taking a suitable Fourier transform) that the non-analytic piece of the amplitude will contribute to the long-range interaction, which is relevant for classical physics. 
\\
\textcolor{black}{
\textbf{Soft expansion and the classical regime:}
Before turning to the detailed computations, we briefly review how the soft expansion emerges and how it encodes the classical limit of the amplitude \cite{Neill:2013wsa,Parra-Martinez:2020dzs}.
For hyperbolic scattering of two compact objects, the classical regime is characterized by an orbital angular momentum that is parametrically large compared to $\hbar$ (Bohr correspondence principle),
\begin{align}
J \gg \hbar \qquad (\text{equivalently, setting }\hbar=1:\; J\gg 1)\, .
\end{align}
Physically, $J\gg 1$ corresponds to a large impact parameter compared to the de Broglie wavelength, so that the scattering process admits a semiclassical description.
In this limit, the momentum transfer $q$ is small compared to the hard external scales set by the Mandelstam invariants and the masses, and classical contributions arise from the non-analytic (long-distance) dependence on $t=q^2$. More precisely, the large-$J$ limit induces a hierarchy of scales,
\begin{align}
s,\; |u|,\; m_i^2 \sim J^2 |t| \gg |t| = |q|^2 \, ,
\end{align}
so that expanding the amplitude at large $J$ is equivalent to expanding in small $|q|$.
This is what we refer to as the \emph{soft expansion}: it is an expansion around small momentum transfer, keeping the hard external kinematics fixed.
Within this expansion, the classical pieces of the amplitude can be systematically isolated (for instance, as the leading terms in $|q|$ at fixed $s$ and masses). It is useful to distinguish three classical limits, depending on how relativistic the scattering is:
\begin{align}
\begin{split}
    &\textrm{Generic classical limit:} \qquad\qquad\qquad\;\; J \gg 1,\\
    &\textrm{Near-static (PN) classical limit:} \qquad\,\,\, J \gg 1,\;\; \sigma \sim 1,\\
    &\textrm{High-energy classical limit:} \qquad\qquad\;\; J \gg 1,\;\; \sigma \gg 1,
\end{split}
\end{align}
where the boost parameter,
\begin{align}
\sigma \equiv \frac{k_1\!\cdot\! k_2}{m_1 m_2}
\end{align}
measures the relative velocity: $\sigma\simeq 1$ corresponds to the near-static/post-Newtonian regime, while $\sigma\gg 1$ captures ultra-relativistic scattering, where we have a stricter scale hierarchy: $s,|u|\gg m_i^2\sim J^2|t|\gg |t|=|q|^2$.
Since our goal is the conservative sector of the dynamics, we further restrict to the \emph{instantaneous} (or potential) long-range kinematics for the momentum transfer,
\begin{align}
q=(q^0,\boldsymbol{q})\,,\qquad |\boldsymbol{q}| \gg q^0\, .
\end{align}
This ensures that the long-range interaction is effectively instantaneous (no on-shell radiation is exchanged), which is precisely the regime relevant for defining a conservative potential from amplitudes. With these scalings in mind, loop integrals are conveniently analyzed using the method of regions.
For an internal graviton (or dilaton) momentum $\ell=(\omega,\boldsymbol{\ell})$, one encounters four standard momentum regions, distinguished by how $(\omega,\boldsymbol{\ell})$ compares to the hard scale $m$ and the soft scale $|\boldsymbol{q}|$:
\begin{align}
\begin{split}
    &\text{hard:}\qquad\qquad\;\; (\omega,\boldsymbol{\ell}) \sim (m,m),\\
    &\text{soft:}\qquad\qquad\;\; (\omega,\boldsymbol{\ell}) \sim (|\boldsymbol{q}|,|\boldsymbol{q}|) 
    \sim J^{-1}(m|\boldsymbol{v}|,\,m|\boldsymbol{v}|),\\
    &\text{potential:}\qquad\;\; (\omega,\boldsymbol{\ell}) \sim (|\boldsymbol{q}|\,|\boldsymbol{v}|,\,|\boldsymbol{q}|)
    \sim J^{-1}(m|\boldsymbol{v}|^2,\,m|\boldsymbol{v}|),\\
    &\text{radiation:}\qquad (\omega,\boldsymbol{\ell}) \sim (|\boldsymbol{q}|\,|\boldsymbol{v}|,\,|\boldsymbol{q}|\,|\boldsymbol{v}|)
    \sim J^{-1}(m|\boldsymbol{v}|^2,\,m|\boldsymbol{v}|^2).
\end{split}
\end{align}
Here we use the reference mass scale $m=m_1+m_2$ and the typical relative velocity scale $|\boldsymbol{v}|$ (which is related to $\sigma$).
The hard region captures short-distance physics and generates local (analytic) contributions, while the soft, potential, and radiation regions encode long-distance physics controlled by the small momentum transfer.
In particular, the \emph{potential} region is the one relevant for conservative dynamics: it corresponds to off-shell exchange with $\omega \ll |\boldsymbol{\ell}|$, matching the instantaneous regime $|\boldsymbol{q}|\gg q^0$.
The \emph{radiation} region, by contrast, is associated with near on-shell modes and is responsible for dissipative effects; we will therefore exclude it when focusing on the conservative potential. Except for the hard region, the remaining regions are governed by two small parameters: the soft scale $|\boldsymbol{q}|\sim J^{-1}$ controlling the classical expansion, and the velocity (equivalently $\sigma$) controlling the non-relativistic/PN/near-static expansion.
This double expansion will be the organizing principle for the computations that follow.}
\\\\
\textbf{Box topologies:}
To compute the amplitude from the box topologies, we encounter the following family of master integrals:
\begin{align}
    G^{\square}_{\alpha_1\alpha_2\alpha_3\alpha_4}=\int \frac{d^D\ell}{(2\pi)^D}\frac{1}{\rho_1^{\alpha_1}\rho_2^{\alpha_2}\rho_3^{\alpha_3}\rho_4^{\alpha_4}}
\end{align}
with, $\rho_1=\ell^2,\,\rho_2=(\ell-q)^2,\,\rho_3=(\ell+k_1)^2-m_1^2, \rho_4=(\ell-k_2)^2-m_2^2$. Now, for convenience, we choose the following kinematics,
\begin{align}
    \begin{split}
        k_1=\bar m_1 u_1-\frac{q}{2}, \,k_2=\bar m_2 u_2+\frac{q}{2}\implies k_1'=\bar m_1 u_1+\frac{q}{2},\,k_2'=\bar m_2 u_2-\frac{q}{2},\,
    \end{split}
\end{align}
Note that, by construction, the momentum transfer $q$ is orthogonal to the velocities $\bar u_is$. We would like to expand the full integral in the  soft region (which is important to take the classical limit), which follows the following hierarchy of scales,
\begin{align}
    |\ell|\sim |q|\ll m_i, \sqrt{s}\to (m_1+m_2)\,.
\end{align}
Now expanding the denominators ($D_2, \,D_4$) in this above mentioned limit we will get,
\begin{align}
\begin{split}
    &\rho_3=\ell^2+2 k_1\cdot \ell+i\epsilon= \ell^2+2 (\bar m_1u_1-\frac{q}{2})\cdot \ell+i\epsilon \sim 2 \bar m_1 u_1\cdot \ell+i\epsilon\,,\\ &
    \rho_4=\ell^2-2 k_2\cdot \ell+i\epsilon= \ell^2-2 (\bar m_2u_2+\frac{q}{2})\cdot \ell+i\epsilon \sim -2 \bar m_2 u_2\cdot \ell+i\epsilon\,.
    \end{split}
\end{align}
We parametrize the velocities as follows,
\begin{align}
    u_1=(1,0,0,0), \,u_2=(\sqrt{1+v^2},0,0,v).
\end{align}
Now in the soft region, the integral family reduces to,
\begin{align}
    \begin{split}
        \mathbfcal{G}^{}
_{\gamma_1\gamma_2\gamma_3\gamma_4}=\int \frac{d^d\ell}{(2\pi)^d}\frac{1}{D_1^{\gamma_1}D_2^{\gamma_2}D_3^{\gamma_3}D_4^{\gamma_4}}
    \end{split}
\end{align}
where the new ISPs (irreducible scalar products) are,
\begin{align}
    D_{i}\equiv \Big(2\bar m_1 u_1\cdot \ell+i\epsilon, -2 \bar m_2u_2\cdot \ell+i\epsilon,\ell^2,(\ell-q)^2\Big)\,.
\end{align}
After solving the IBP identities, one will get the following master integrals:
\begin{align}
    \vec f(x,\epsilon)=\{\mathbfcal{G}_{0,0,1,1},\mathbfcal{G}_{0,1,1,1},\mathbfcal{G}_{1,0,1,1},\mathbfcal{G}_{1,1,1,1}\}
\end{align}
where, $x=\sigma-\sqrt{\sigma^2-1}$ with, \textcolor{black}{$\sigma=\frac{k_1\cdot k_2}{m_1m_2}$}. We solve the master integrals by solving the following differential equation:
\begin{align}
    \partial_{x}\vec f(x,\epsilon)=A(x,\epsilon)\vec f(x,\epsilon)
\end{align}
where, the matrix $A(x,\epsilon)$ (using \cite{Gituliar:2017vzm}) is given by (only scale is $|q|$, so we set $q^2\to -1$, which can be restored from dimensional analysis),
\begin{align}
    \begin{split}
        A(x,\epsilon)=\left(
\begin{array}{ccc}
 0 & 0 & 0 \\
 0 & 0 & 0 \\
 \frac{2 (1-2 \epsilon )}{(x-1) (x+1)} & 0 & \frac{-x^2-1}{(x-1) x (x+1)} \\
\end{array}
\right)\,.
    \end{split}
\end{align}
The main goal is to find the canonical basis $ \vec g(x,\epsilon)=\mathbb{T}^{-1}(x,\epsilon)\vec f(x,\epsilon)$ such that the new differential equation will be $\epsilon$-factorized,   
\begin{align}
    \begin{split}
\partial_{x} \vec g(x,\epsilon)=\mathbb{T}^{-1}(A \mathbb{T}-\partial_x \mathbb{T})\vec g(x,\epsilon)\equiv \epsilon\, \mathbb{S}(x) \vec g(x,\epsilon)\,.
    \end{split}
\end{align}
The transformation matrix $\mathbb{T}$ and the $\epsilon$-factorized matrix $\mathbb{S}$ are given by,
\begin{align}
    \begin{split}
        \mathbb{T}(x,\epsilon)=\left(
\begin{array}{ccc}
 \frac{9 \epsilon }{2 \epsilon -1} & 0 & 0 \\
 2 & 1 & 0 \\
 \frac{2 x}{x^2-1} & -\frac{x}{x^2-1} & -\frac{3 x}{x^2-1} \\
\end{array}
\right),\,\,\,\mathbb{S}(x)=\left(
\begin{array}{ccc}
 0 & 0 & 0 \\
 0 & 0 & 0 \\
 \frac{6  }{x} & 0 & 0 \\
\end{array}
\right)\,.
    \end{split}
\end{align}
Therefore, the differential equation becomes,
\begin{align}
  &  d\,\vec{g}(x,\epsilon)=\epsilon\, \widetilde {\mathbb{S}}\,d\log(x)\,\vec g(x,\epsilon),\, \widetilde {\mathbb{S}}=\left(
\begin{array}{ccc}
 0 & 0 & 0 \\
 0 & 0 & 0 \\
 6 & 0 & 0 \\
\end{array}\right)\,.
\end{align}
  Finally, in  the Laporta basis, the formal solution takes the form,
\begin{align}
    \begin{split}
        \vec f(x,\epsilon)=\mathbb{T}(x,\epsilon)\,\Bigg(\mathbfcal{P}e^{\epsilon\int_{1^{-}}^x \,\mathbb{S}(x')\,dx'}\Bigg)\,\mathbb{T}^{-1}(1^{-},\epsilon) \,\vec f(1^{-},\epsilon)\,.\label{3.16}
    \end{split}
\end{align}
Schematically, any one-loop soft-box diagram can be written as,
\begin{equation}
\hspace{-10 cm}\mathbfcal{A}_{\textrm{soft}}^{\textrm{box}}=
   \begin{minipage}
         [h]{0.15\linewidth}
	\vspace{1pt}
	\scalebox{5}{\includegraphics[width=\linewidth]{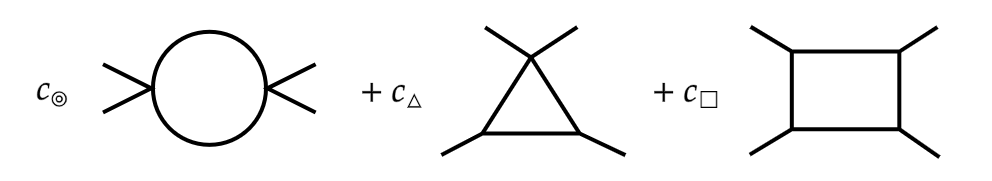}}
    \end{minipage}
  \end{equation}
  To solve the master integrals, we need boundary condition which we take for $x\to 1$ i.e the non relativistic limit.\\\\\\
  \textbf{Re-expansion in static/non-relativistic limit:}
  Now, to fix the boundary condition, we need to re-expand the integrals in the potential region in the near static limit ($\boldsymbol{v}\to 0$). The scaling rules for near static are,
  \begin{align}
      \begin{split}
          u_i^\mu=(u_i^0,\boldsymbol{u}_i)\sim (1,|\boldsymbol v|),\,\ell^\mu=(\omega,\boldsymbol{\ell})\sim |\boldsymbol{q}|(|\boldsymbol{v}|,1).
      \end{split}
  \end{align}
  Now we expand the soft-master integrals in the near static limit,
  \begin{align}
      \begin{split}
         \mathbfcal{G}^{\square}_{1,1,1,1}\xrightarrow[]{\boldsymbol{v}\to 0}\int_{-\infty}^{\infty} \frac{d\omega}{2\pi}\int \frac{d^{d-1}\ell}{(2\pi)^{d-1}} \frac{1}{ (2\bar m_1 \omega+i\epsilon)(-2\bar m_2 u_2^0\omega+2\bar m_2\boldsymbol{u_2}\cdot \boldsymbol{\ell}+i\epsilon)(\boldsymbol{-\ell^2}+i\epsilon)(\boldsymbol{-(\ell-q)^2}+i\epsilon)}\,.
      \end{split}
  \end{align}
Now integrating over $\omega$ (closing the contour in upper/lower half of $\omega$ plane) we get,
\begin{align}
    \begin{split}
       \hspace{-1.9 cm}\mathbfcal{G}^{\square}_{1,1,1,1}\xrightarrow[]{\boldsymbol{v}\to 0}&-\frac{i}{8\pi\bar m_1\bar m_2}\int \frac{d^d\boldsymbol{\ell}}{(2\pi)^{d-1}}\frac{1}{(\boldsymbol{u}_2\cdot \boldsymbol{\ell}+i\epsilon)(\boldsymbol{\ell}^2-i\epsilon)((\boldsymbol{\ell}-\boldsymbol{q})^2-i\epsilon)}\,,\\ &
        \to -\frac{1}{16\pi\bar m_1 \bar m_2}\int \frac{d^{d-1}\ell}{(2\pi)^{d-1}}\frac{\hat\delta(\boldsymbol{u}_2\cdot \boldsymbol{\ell})}{(\boldsymbol{\ell}^2-i\epsilon)((\boldsymbol{\ell-q})^2-i\epsilon)}\,,\\ &
        =-\frac{4^{\epsilon -2} \pi ^{\epsilon } \left(-q^2\right)^{-\epsilon -1} \Gamma (-\epsilon )^2 \Gamma (\epsilon +1)}{2\pi \bar{m}_1 \bar{m}_2 \Gamma (-2 \epsilon )}\,.
    \end{split}
\end{align}
Similarly, the near static limit of  $\mathbfcal{G}_{0,1,1,1}$ evaluates to,
\begin{align}
    \begin{split}
        \mathbfcal{G}^{\square}_{0,1,1,1}&\xrightarrow[]{v\to 0}-\frac{1}{4\pi\bar m_2}\int \frac{d^{d-1}\ell}{(2\pi)^{d-1}} \frac{1}{(\boldsymbol{\ell}^2-i\epsilon)[(\boldsymbol{\ell-\boldsymbol{q}})^2-i\epsilon]}\int_{-\infty}^{\infty}  \frac{d\omega}{\omega-\boldsymbol{u}_2\cdot \boldsymbol{\ell}-i\epsilon}\,,\\ &
        =-\frac{i\pi}{4\pi\bar m_2}\int 
    \frac{d^{d-1}\boldsymbol{\ell}}{(2\pi)^{d-1}}\frac{1}{(\boldsymbol{\ell}^2-i\epsilon)[(\boldsymbol{\ell-\boldsymbol{q}})^2-i\epsilon]}\,,\\ &
        =-\frac{i 16^{\epsilon -1} \pi ^{\epsilon +1} \left(-q^2\right)^{-\epsilon -\frac{1}{2}} \sec (\pi  \epsilon )}{2\pi\bar{m}_2 \Gamma (1-\epsilon )}
    \end{split}
\end{align}
and,
\begin{align}
    \begin{split}
        \mathbfcal{G}_{0,0,1,1}\xrightarrow[]{v\to 0}0\,.
    \end{split}
\end{align}
Now using (\ref{3.16}), the master integrals in the Laporta basis are given by,

\begin{center}
\begin{tcolorbox}[enhanced,  height=6 cm, width=12 cm, colback=brown!2!white, colframe=black!2000!brown, title=\textit{Master Integrals for box topologies}, breakable]

\begin{align}
    \begin{split}
     \hspace{0 cm}    &\mathbfcal{G}^{\square}_{0,0,1,1}(x,\epsilon)=0\,,\\&
        \mathbfcal{G}^{\square}_{0,1,1,1}(x,\epsilon)=- \textcolor{black}{\frac{1}{2 \pi}} \frac{i 16^{\epsilon -1} \pi ^{\epsilon +1} \sec (\pi  \epsilon )}{\bar{m}_2 \Gamma (1-\epsilon )}(-q^2)^{-\epsilon-\frac{1}{2}}\,,\\ &
        \mathbfcal{G}^{\square}_{1,0,1,1}(x,\epsilon)=- \textcolor{black}{\frac{1}{2 \pi}} \frac{i 16^{\epsilon -1} \pi ^{\epsilon +1} \sec (\pi  \epsilon )}{\bar{m}_1 \Gamma (1-\epsilon )}(-q^2)^{-\epsilon-\frac{1}{2}}\,,
        \\ &\mathbfcal{G}^{\square}_{1,1,1,1}(x,\epsilon)= \textcolor{black}{\frac{1}{2 \pi}} \frac{x\, 4^{2 \epsilon -1} \pi ^{\epsilon +\frac{3}{2}} \csc (\pi  \epsilon )}{\left(1-x^2\right) \bar{m}_1 \bar{m}_2 \Gamma \left(\frac{1}{2}-\epsilon \right)}(-q^2)^{-\epsilon-1}\,. 
    \end{split}
\end{align}
\end{tcolorbox}
\end{center}
\section*{Crossed-box topologies:}
In the crossed-box topologies, we encounter the following family of master integrals,
\begin{align}
    \begin{split}
        G^{\boxtimes}_{\gamma_1\gamma_2\gamma_3\gamma_4}=\int \frac{d^d\ell}{(2\pi)^d} \frac{1}{{\rho_1}^{\gamma_1}{\rho_2}^{\gamma_2}{\rho_4}^{\gamma_3}\tilde{\rho_4}^{\gamma_4}},
    \end{split}
\end{align}
with,  $\rho_1=\ell^2,\,\rho_2=(\ell-q)^2,\,\rho_3=(\ell+k_1)^2-m_1^2, \tilde\rho_4=(\ell+k_2-q)^2-m_2^2$. As mentioned earlier, we are interested in the soft/potential region of the amplitude. In the soft limit, the family reduces to,
\begin{align}
     \mathbfcal{G}^{\boxtimes}
_{\gamma_1\gamma_2\gamma_3\gamma_4}=\int \frac{d^d\ell}{(2\pi)^d}\frac{1}{\tilde D_1^{\gamma_1}\tilde D_2^{\gamma_2}\tilde D_3^{\gamma_3}\tilde D_4^{\gamma_4}}
\end{align}
where, 
$\tilde D_1=2\ell\cdot u_1+i\epsilon, \,\tilde D_2=2\ell\cdot u_2+i\epsilon,\, \tilde D_3=\ell^2,\,\tilde D_4=(\ell-q)^2. $ As we see in the crossed box topology, the two linearized matter propagators have the same sign, unlike the box integrals. As with the box topology, in this case also we have three master integrals: $\mathcal{G}_{0,0,1,1}^{\boxtimes},\,\mathcal{G}_{0,1,1,1}^{\boxtimes},\,\mathcal{G}_{1,1,1,1}^{\boxtimes}\,.$ Now, one will take a similar procedure as before to solve the integrals. However, one would notice that in the near static limit, the last will be zero.
\begin{align}\label{3.28} \mathbfcal{G}_{1,1,1,1}^{\boxtimes}=\frac{1}{4\bar{m}_1\bar{m}_2}\int \frac{d^{d-1}\boldsymbol{\ell}}{(2\pi)^{d-1}}\frac{1}{(\boldsymbol{\ell}^2-i\epsilon)\left[(\boldsymbol{\ell}-\boldsymbol{q})^2-i\epsilon\right]}\int \frac{d\omega}{2\pi} \frac{1}{(\omega+i\epsilon)(\omega+\boldsymbol{u}_2\cdot\boldsymbol{\ell}+i\epsilon)}\,.
\end{align}
The $\omega$ integral can be shown to be zero by closing the contour in the upper or lower half plane and computing the residues.
\begin{center}
\begin{tcolorbox}[enhanced,  height= 5 cm, width=12 cm, colback=brown!2!white, colframe=black!2000!brown, title=\textit{Master Integrals for crossed-box topologies}, breakable]
\begin{centering}
\begin{align}
    \begin{split}
     \hspace{0 cm}    &\mathbfcal{G}^{\boxtimes}_{0,0,1,1}(x,\epsilon)=0\,,\\&
        \mathbfcal{G}^{\boxtimes}_{0,1,1,1}(x,\epsilon)=- \textcolor{black}{\frac{1}{2 \pi}} \frac{i 16^{\epsilon -1} \pi ^{\epsilon +1} \sec (\pi  \epsilon )}{\bar{m}_2 \Gamma (1-\epsilon )}(-q^2)^{-\epsilon-\frac{1}{2}}\,\\ &
        \mathbfcal{G}^{\boxtimes}_{1,0,1,1}(x,\epsilon)=- \textcolor{black}{\frac{1}{2 \pi}} \frac{i 16^{\epsilon -1} \pi ^{\epsilon +1} \sec (\pi  \epsilon )}{\bar{m}_1 \Gamma (1-\epsilon )}(-q^2)^{-\epsilon-\frac{1}{2}}\,\\&\mathbfcal{G}^{\boxtimes}_{1,1,1,1}(x,\epsilon)=0\,. 
    \end{split}
\end{align}
\end{centering}
\end{tcolorbox}
\end{center}
\noindent
Now we are ready to evaluate all the diagrams that contribute to the one-loop scattering amplitude in the soft limit. Below, we give the computational details. Before going to the one-loop computation we first give the results for tree level amplitude as it will play a crucial role to define IR finite potential.\\\\
\textbf{Tree-amplitude:}\\
\hspace{-3 cm}
\begin{align}
\begin{split}
&i\mathbfcal{A}_{4}^{\texttt{tree}}\left[\begin{minipage}
          [h]{0.40\linewidth}
	\vspace{1.2 pt}
	\scalebox{1.5}{\includegraphics[width=\linewidth]{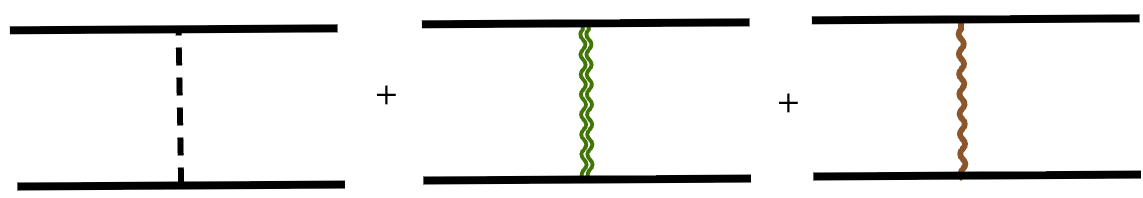}}
      \end{minipage}\hspace{3.2 cm}\right]\\ &=\frac{1}{q^2}\left(-\frac{4  a^2 m_1 m_2}{m_p^2}+16  \alpha _1 \alpha _2 \texttt{g}_ce^2 m_1 m_2 \sigma  +\frac{ m_1^2 m_2^2 \left(-2 \sigma ^2 +1\right)}{2 m_p^2}\right)
    +\left(4  \alpha _1 \alpha _2 \texttt{g}_c e^2 -\frac{ 2 m_2 m_1 \sigma   }{4 m_p^2}\right)\,.    \end{split}
\end{align}
\noindent
We now move on to the computation of one-loop amplitudes in the soft limit and, consequently, use them to compute the classical Post-Minkowskian potential and the conservative scattering angle. \\\\
\textbf{One-loop amplitudes: }\\\\
In this section, we derive the one-loop amplitudes for different topologies and give the necessary details.  Expressions for the numerators of all the Feynman diagrams are given in the ancillary file \texttt{numerators\_emd.nb} submitted to the \texttt{arXiv} along with this paper. 
We start with computing the box (+ cross-box) amplitude.\\\\
\textbf{Soft amplitudes from Box (cross box) topologies:}\\
\textbullet $\,\,$ The amplitude for the box (+cross-box) diagram with double photon exchange is given by,

\begin{flalign}
& i\mathbfcal{A}_{4,\textrm{soft}}^{1-\textrm{loop}}\left[\begin{minipage}
          [h]{0.40\linewidth}
	\vspace{1.2 pt}
	\scalebox{1}{\includegraphics[width=\linewidth]{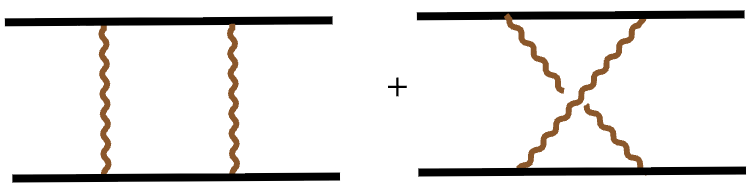}}
      \end{minipage}\right]=\int_{\ell}\frac{\mathcal{N}^{(1)}_{\filledsquare{gray}}}{\rho_1 \rho_2 \rho_3 \rho_4}\Bigg|_{\textrm{soft}} +\int_{\ell}\frac{\mathcal{N}^{(1)}_{\Join}}{\rho_1 \rho_2 \rho_3 \tilde\rho_4}\Bigg|_{\textrm{soft}} \nonumber \\
      & = \frac{\alpha _1^2 \alpha _2^2 e^4 m_1 m_2 \sigma ^2 16^{\epsilon +1} \pi ^{\epsilon +\frac{1}{2}} \texttt{g}_c^2 \left(-q^2\right)^{-\epsilon -1} \csc (\pi  \epsilon )}{\sqrt{\sigma ^2-1} \Gamma \left(\frac{1}{2}-\epsilon \right)}\,. &
\end{flalign}
\\ 
\textbullet $\,\,$Amplitude corresponding to the box diagram having photon and graviton exchange is given by,      

      \begin{align}
\begin{split}
&i\mathbfcal{A}_{4,\textrm{soft}}^{1-\textrm{loop}}\left[\begin{minipage}
          [h]{0.40\linewidth}
	\vspace{1.2 pt}
	\scalebox{1}{\includegraphics[width=\linewidth]{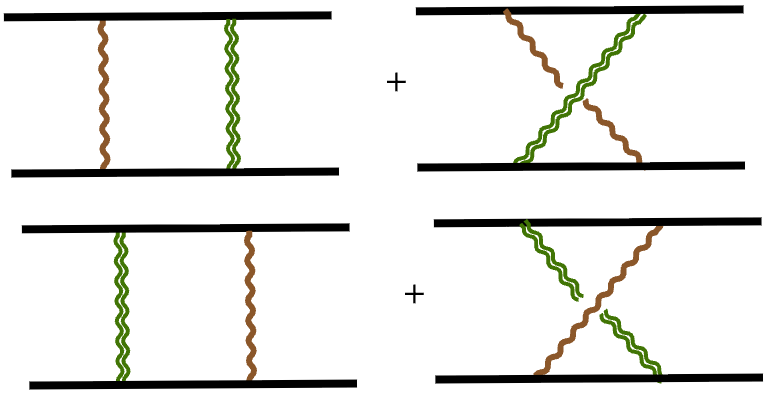}}
      \end{minipage}\right]=\sum_{i=a,c}\int_{\ell}\frac{\mathcal{N}^{(2i)}_{\filledsquare{gray}}}{\rho_1 \rho_2 \rho_3 \rho_4}\Bigg|_{\textrm{soft}}+\sum_{j=b,d}\int_{\ell}\frac{\mathcal{N}^{(2j)}_{\Join}}{\rho_1 \rho_2 \rho_3 \tilde\rho_4}\Bigg|_{\textrm{soft}}\\ &
      =    \alpha _1 \alpha _2 e^2 m_1 m_2 2^{4 \epsilon -3} \pi ^{\epsilon } \texttt{g}_c \left(-q^2\right)^{-\epsilon } \left(\frac{-8 \sqrt{\pi } m_1 m_2 \sigma  \left(2 \sigma ^2 (\epsilon -1)+1\right) \csc (\pi  \epsilon )}{-q^2 \sqrt{\sigma ^2-1} (\epsilon -1) m_p^2 \Gamma \left(\frac{1}{2}-\epsilon \right)} \right.\\
      & \left. -\frac{i \left(m_1+m_2\right) \left(6 \sigma ^2 (\epsilon -1)+2 \sigma  \epsilon +1\right) \sec (\pi  \epsilon )}{\sqrt{-q^2} (\epsilon -1) m_p^2 \Gamma (1-\epsilon )}-\frac{i \left(m_1+m_2\right) \left(6 \sigma ^2 (\epsilon -1)-2 \sigma  \epsilon +1\right) \sec (\pi  \epsilon )}{\sqrt{-q^2} m_p^2 \Gamma (2-\epsilon )}\right).
      \end{split}
      \end{align}
Note that, a and b denotes the diagrams of the first row diagram in the first row and c and d denotes the diagrams of the second row.\\\\       
\textbullet $\,\,$The amplitude corresponding to the  box (+ cross-box) diagram with dilaton and photon exchange is given by,

 \begin{flalign}
& i\mathbfcal{A}_{4,\textrm{soft}}^{1-\textrm{loop}}\left[\begin{minipage}
          [h]{0.40\linewidth}
	\vspace{1.2 pt}
	\scalebox{1}{\includegraphics[width=\linewidth]{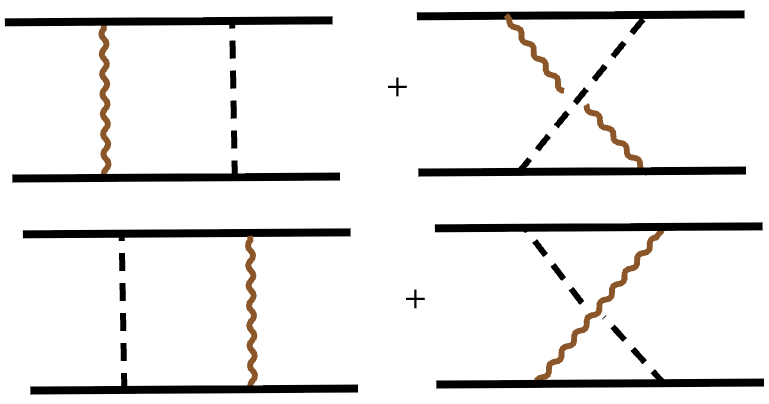}}
      \end{minipage}\right] =\sum_{i=a,c}\int_{\ell}\frac{\mathcal{N}^{(3i)}_{\filledsquare{gray}}}{\rho_1 \rho_2 \rho_3 \rho_4}\Bigg|_{\textrm{soft}}+\sum_{j=b,d}\int_{\ell}\frac{\mathcal{N}^{(3j)}_{\Join}}{\rho_1 \rho_2 \rho_3 \tilde\rho_4}\Bigg|_{\textrm{soft}} \nonumber \\
      & = -\frac{a^2 \alpha _1 \alpha _2 e^2 m_1 m_2 \sigma  2^{4 \epsilon +3} \pi ^{\epsilon +\frac{1}{2}} \texttt{g}_c \left(-q^2\right)^{-\epsilon -1} \csc (\pi  \epsilon )}{\sqrt{\sigma ^2-1} m_p^2 \Gamma \left(\frac{1}{2}-\epsilon \right)} . &
      \end{flalign}
      \newpage
\textbullet $\,\,$ The amplitude of the box (+cross-box) diagram corresponding to the double dilaton exchange is given by,
\begin{align}
\begin{split}
&i\mathbfcal{A}_{4,\textrm{soft}}^{1-\textrm{loop}}\left[\begin{minipage}
          [h]{0.40\linewidth}
	\vspace{1.2 pt}
	\scalebox{1}{\includegraphics[width=\linewidth]{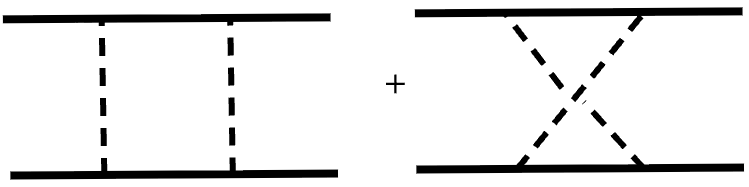}}
    \end{minipage}\right] =\int_{\ell}\frac{\mathcal{N}^{(4)}_{\filledsquare{gray}}}{\rho_1 \rho_2 \rho_3 \rho_4}\Bigg|_{\textrm{soft}}+\int_{\ell}\frac{\mathcal{N}^{(4)}_{\Join}}{\rho_1 \rho_2 \rho_3 \tilde\rho_4}\Bigg|_{\textrm{soft}} \\ &= \frac{a^4 2^{4 \epsilon -3} \pi ^{\epsilon +\frac{1}{2}} \left(-q^2\right)^{-\epsilon -1} \csc (\pi  \epsilon ) \left(8 m_1^2 m_2^2 \left(\sigma ^2-1\right)-q^2 \left(2 m_2 m_1 \sigma +m_1^2+m_2^2\right)\right)}{m_1 m_2 \left(\sigma ^2-1\right)^{3/2} m_p^4 \Gamma \left(\frac{1}{2}-\epsilon \right)}\,.
      \end{split}
      \end{align}
      \textbullet $\,\,$ The amplitude of the box (+cross-box) diagram corresponding to the graviton-dilaton  exchange is given by,
\begin{flalign}
& i\mathbfcal{A}_{4,\textrm{soft}}^{1-\textrm{loop}}\left[\begin{minipage}
          [h]{0.40\linewidth}
	\vspace{1.2 pt}
	\scalebox{1}{\includegraphics[width=\linewidth]{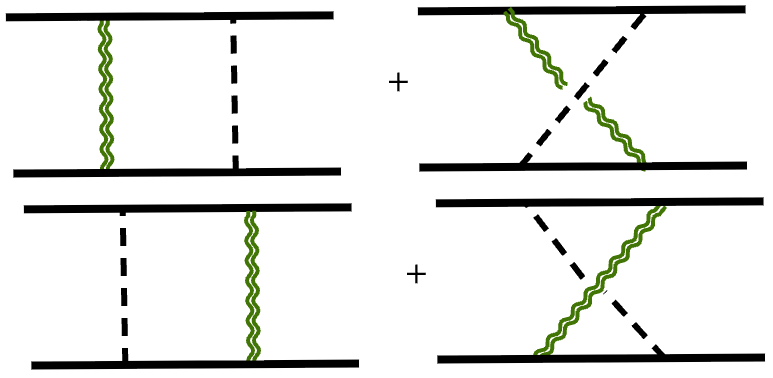}}
    \end{minipage}\right]    =\int_{\ell}\frac{\mathcal{N}^{(5)}_{\filledsquare{gray}}}{\rho_1 \rho_2 \rho_3 \rho_4}\Bigg|_{\textrm{soft}}+\int_{\ell}\frac{\mathcal{N}^{(5)}_{\Join}}{\rho_1 \rho_2 \rho_3 \tilde\rho_4}\Bigg|_{\textrm{soft}}\nonumber\\ &
    =\frac{a^2 m_1^2 m_2^2 2^{4 \epsilon -2} \pi ^{\epsilon +\frac{1}{2}} \left(-q^2\right)^{-\epsilon -1} \left(-2 \sigma ^2+2 \sigma ^2 \epsilon +1\right) \csc (\pi  \epsilon )}{\sqrt{\sigma ^2-1} (\epsilon -1) m_p^4 \Gamma \left(\frac{1}{2}-\epsilon \right)}  \nonumber \\
    & \hspace{0 cm} -\frac{i a^2 m_1 m_2 \left(m_1+m_2\right) \sigma  4^{2 \epsilon -1} \pi ^{\epsilon } \left(-q^2\right)^{-\epsilon -\frac{1}{2}} \sec (\pi  \epsilon )}{m_p^4 \Gamma (1-\epsilon )}\,. &
\end{flalign}
\textbullet $\,\,$   The amplitude of the box (+cross-box) diagram corresponding to the graviton-graviton  exchange is given by,  
\begin{flalign}
&\hspace{-0 
cm}i\mathbfcal{A}_{4,\textrm{soft}}^{1-\textrm{loop}}\left[\begin{minipage}
          [h]{0.40\linewidth}
	\vspace{1.2 pt}
	\scalebox{1}{\includegraphics[width=\linewidth]{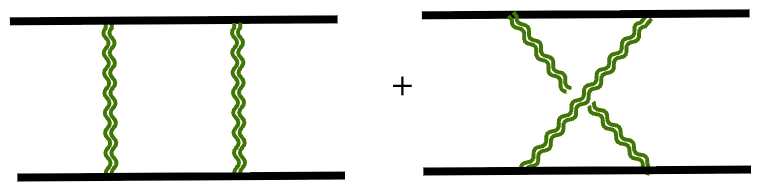}}
    \end{minipage}\right]    =\int_{\ell}\frac{\mathcal{N}^{(6)}_{\filledsquare{gray}}}{\rho_1 \rho_2 \rho_3 \rho_4}\Bigg|_{\textrm{soft}}+\int_{\ell}\frac{\mathcal{N}^{(6)}_{\Join}}{\rho_1 \rho_2 \rho_3 \tilde\rho_4}\Bigg|_{\textrm{soft}} \nonumber \\ & =\frac{m_1 m_2 2^{4 \epsilon -9} \pi ^{\epsilon +\frac{1}{2}} \left(-q^2\right)^{-\epsilon }  \csc (\pi  \epsilon ) \left(8 m_1^2 m_2^2 \left(\sigma ^2-1\right)-q^2 \left(2 m_2 m_1 \sigma +m_1^2+m_2^2\right)\right)}{-q^2 \left(2 \sigma ^2 (\epsilon -1)+1\right)^{-2} \left(\sigma ^2-1\right)^{3/2} (\epsilon -1)^2 m_p^4 \Gamma \left(\frac{1}{2}-\epsilon \right)} \nonumber \\ 
    & \hspace{1cm}+\frac{i m_1^2 m_2^2 \left(m_1+m_2\right) 4^{2 \epsilon -3} \pi ^{\epsilon } \epsilon  \left(-q^2\right)^{-\epsilon -\frac{1}{2}} \left(2 \sigma ^2 (\epsilon -1)+1\right) \sec (\pi  \epsilon )}{(\epsilon -1)^2 m_p^4 \Gamma (1-\epsilon )}\,.  &  
\end{flalign}
As one can immediately see by expanding the box(+cross-box) amplitudes, the $\frac{1}{\sqrt{-q^2}}$ terms cancelled, so they do not contribute to the classical potential. \newpage  
\subsection*{Soft amplitudes from triangle topologies:}   
\textbullet  $\,\,$The soft amplitude corresponding to the triangle topology with photon lines is given by,


\begin{flalign}
&\hspace{-0 cm}i\mathbfcal{A}_{4,\textrm{soft}}^{1-\textrm{loop}}\left[\begin{minipage}
          [h]{0.40\linewidth}
	\vspace{1.2 pt}
	\scalebox{0.8}{\includegraphics[width=\linewidth]{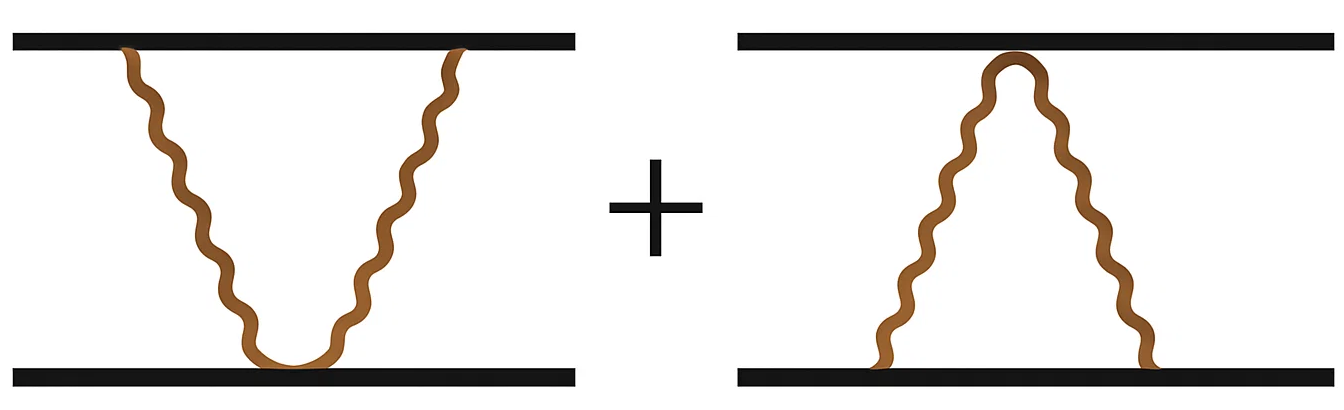}}
    \end{minipage}\hspace{- 1.3 cm}\right]=\int_{\ell}\frac{\mathcal{N}^{(1a)}_{\triangle}\rho_4+\mathcal{N}^{(1b)}_{\triangle}\rho_3}{\rho_1 \rho_2 \rho_3 \rho_4}\Bigg|_{\textrm{soft}} \nonumber \\ 
    &\hspace{0 cm} =\frac{i \alpha _1^2 \alpha _2^2 e^4 \left(m_1+m_2\right) 4^{2 \epsilon +1} \pi ^{\epsilon } g_s^2 \left(-q^2\right)^{-\epsilon -\frac{1}{2}} \sec (\pi  \epsilon )}{\Gamma (1-\epsilon )} \,. &
\end{flalign}\\

\textbullet $\,\,$ The soft amplitude corresponding to the triangle topology with one graviton and one photon line is given by,   
\begin{align}
\begin{split}
&\hspace{-0 cm}i\mathbfcal{A}_{4,\textrm{soft}}^{1-\textrm{loop}}\left[\begin{minipage}
          [h]{0.47\linewidth}
	\vspace{1.2 pt}
	\scalebox{0.8}{\includegraphics[width=\linewidth]{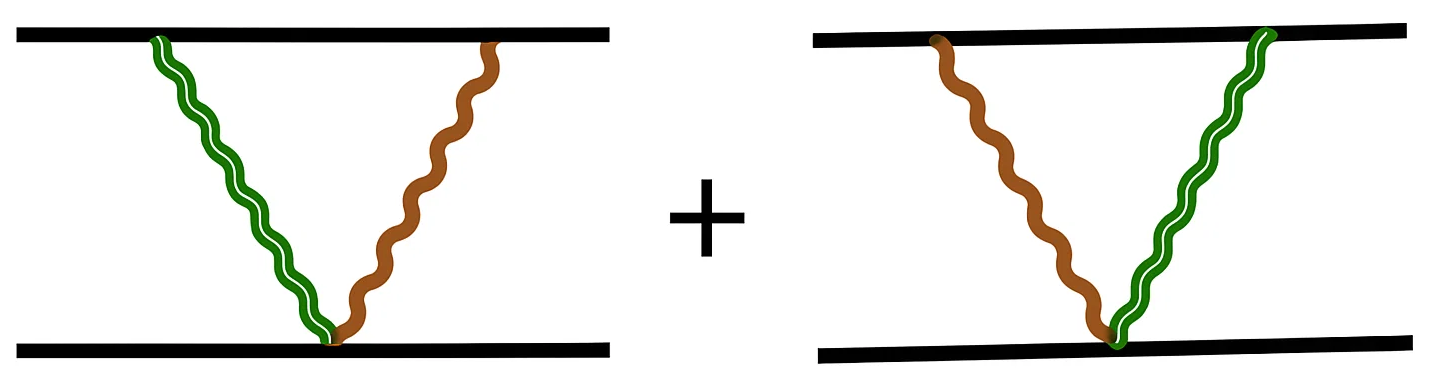}}
    \end{minipage}\hspace{-1.5 cm}\right] =\int_{\ell}\frac{\mathcal{N}^{(2a)}_{\triangle}}{\rho_1 \rho_2 \rho_3 } \Bigg|_{\textrm{soft}} \\ &\hspace{-0 cm}
    =-\frac{i \alpha _1 \alpha _2 e^2 2^{4 \epsilon -3} \pi ^{\epsilon } g_s \left(-q^2\right)^{-\epsilon -\frac{1}{2}} \sec (\pi  \epsilon )}{(\epsilon -1)^2 m_p^2 \Gamma (1-\epsilon )}\left[m_1 q^2 \left(\epsilon ^2-\epsilon +1\right)+m_2 q^2 \sigma  \epsilon +8 m_1^2 m_2 \sigma  \left(\epsilon ^2-\epsilon +1\right)\right]
      \end{split}
      \end{align}
and,
\begin{align}
\begin{split}
&\hspace{0 cm}i\mathbfcal{A}_{4,\textrm{soft}}^{1-\textrm{loop}}\left[\begin{minipage}
          [h]{0.47\linewidth}
	\vspace{1.2 pt}
	\scalebox{0.9}{\includegraphics[width=\linewidth]{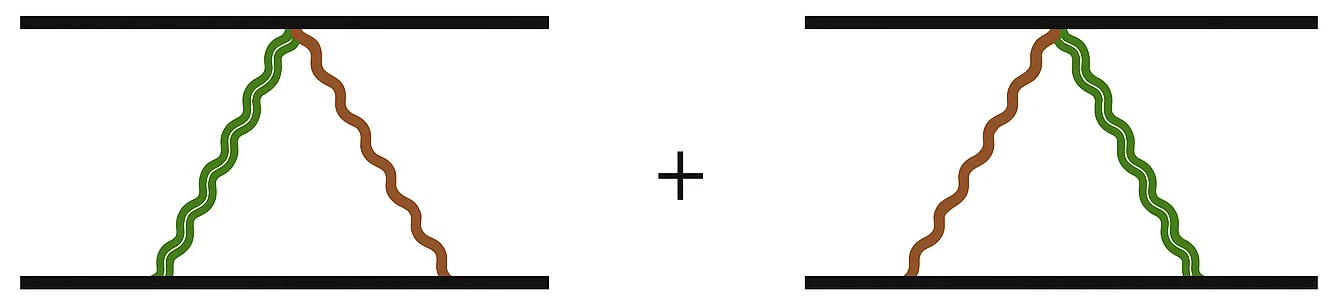}}
    \end{minipage}\hspace{-0.5 cm}\right]=  \int_{\ell}\frac{\mathcal{N}^{(2b)}_{\triangle}}{\rho_1 \rho_2 \rho_4 } \Bigg|_{\textrm{soft}} \\ &\hspace{-0 cm} =-\frac{i \alpha _1 \alpha _2 e^2 2^{4 \epsilon -3} \pi ^{\epsilon } g_s \left(-q^2\right)^{-\epsilon -\frac{1}{2}} \sec (\pi  \epsilon )}{(\epsilon -1)^2 m_p^2 \Gamma (1-\epsilon )}\left[m_1 \sigma  \left(8 m_2^2 \left(\epsilon ^2-\epsilon +1\right)+q^2 \epsilon \right)+m_2 q^2 \left(\epsilon ^2-\epsilon +1\right)\right]\,.
      \end{split}
      \end{align}\\
\textbullet $\,\,$ The soft amplitude corresponding to the triangle topology with graviton lines are given by,   
\begin{align}
\begin{split}
&\hspace{-0 cm}i\mathbfcal{A}_{4,\textrm{soft}}^{1-\textrm{loop}}\left[\begin{minipage}
          [h]{0.50\linewidth}
	\vspace{1.2 pt}
	\scalebox{0.8}{\includegraphics[width=\linewidth]{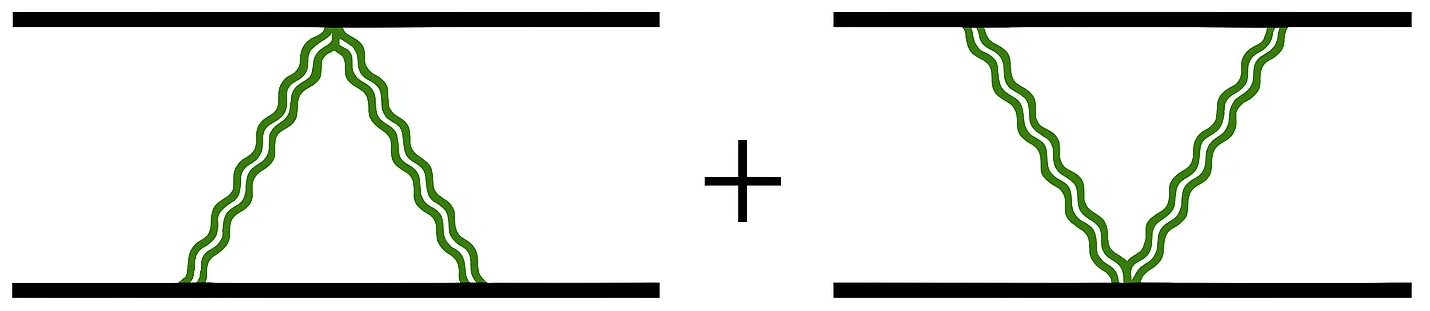}}
\end{minipage}\hspace{-1 cm}\right]=\int_{\ell}\frac{\mathcal{N}^{(3a)}_{\triangle}\rho_4+\mathcal{N}^{(3b)}_{\triangle}\rho_3}{\rho_1 \rho_2 \rho_3 } \Bigg|_{\textrm{soft}}   \\ &\hspace{-0.3 cm} =\frac{i m_1^2 m_2^2 \left(m_1+m_2\right) 2^{4 \epsilon -6} \pi ^{\epsilon } \left(-q^2\right)^{-\epsilon -\frac{1}{2}} \left(2 \sigma ^2 (\epsilon -1)^2-\epsilon \right) \sec (\pi  \epsilon )}{(\epsilon -1)^2 m_p^4 \Gamma (1-\epsilon )}\,.
      \end{split}
      \end{align}  
\textbullet $\,\,$ The soft amplitude corresponding to the triangle topology with dilaton lines is given by,   
\begin{align}
\begin{split}
&\hspace{-0 cm}i\mathbfcal{A}_{4,\textrm{soft}}^{1-\textrm{loop}}\left[\begin{minipage}
          [h]{0.50\linewidth}
	\vspace{1.2 pt}
	\scalebox{0.9}{\includegraphics[width=\linewidth]{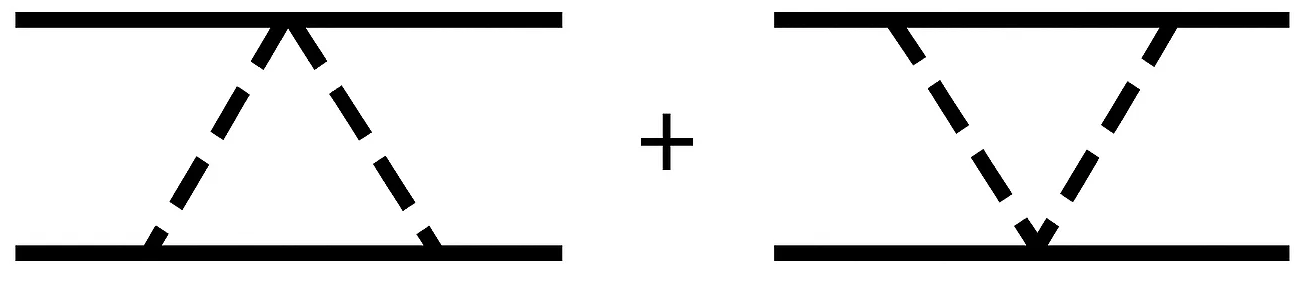}}
    \end{minipage}\hspace{-0.8 cm}\right] =\int_{\ell}\frac{\mathcal{N}^{(4a)}_{\triangle}\rho_4+\mathcal{N}^{(4b)}_{\triangle}\rho_3}{\rho_1 \rho_2 \rho_3 \rho_4 } \Bigg|_{\textrm{soft}}   \\ &\hspace{-0.3 cm} =\frac{i a^2 2^{4 \epsilon -5} \pi ^{\epsilon } \left(-q^2\right)^{-\epsilon -\frac{1}{2}} \sec (\pi  \epsilon ) \left(8 a^2 m_1 m_2 \left(m_1+m_2\right)-32 b m_1^2 m_2^2\right)}{m_1 m_2 m_p^4 \Gamma (1-\epsilon )}\,.
      \end{split}
      \end{align}\\
      \textbullet $\,\,$ The soft amplitude corresponding to the triangle topology with one dilaton line and one graviton line is given by,   

\begin{flalign}
& i\mathbfcal{A}_{4,\textrm{soft}}^{1-\textrm{loop}}\left[\begin{minipage}
          [h]{0.50\linewidth}
	\vspace{1.2 pt}
	\scalebox{0.8}{\includegraphics[width=\linewidth]{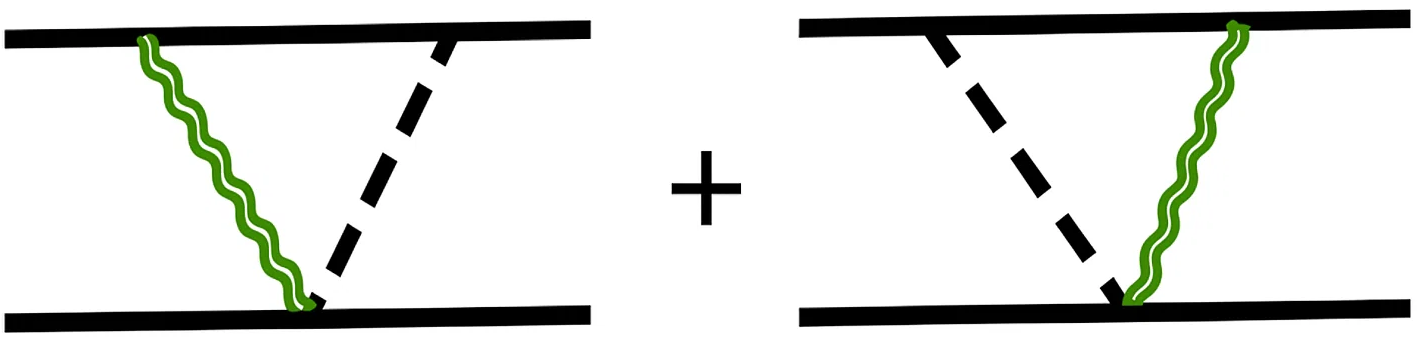}}
    \end{minipage}\hspace{-1 cm}\right]=\int_{\ell}\frac{\mathcal{N}^{(5a)}_{\triangle}}{\rho_1 \rho_2 \rho_3 } \Bigg|_{\textrm{soft}} \nonumber   \\
    &  =-\frac{i a^2 m_2 4^{2 \epsilon -3} \pi ^{\epsilon } (\epsilon +1) \left(8 m_1^2+q^2\right) \left(-q^2\right)^{-\epsilon -\frac{1}{2}} \sec (\pi  \epsilon )}{(\epsilon -1) m_p^4 \Gamma (1-\epsilon )}, &
\end{flalign} 

and,

\begin{flalign}
& i\mathbfcal{A}_{4,\textrm{soft}}^{1-\textrm{loop}}\left[\begin{minipage}
          [h]{0.50\linewidth}
	\vspace{1.2 pt}
	\scalebox{0.8}{\includegraphics[width=\linewidth]{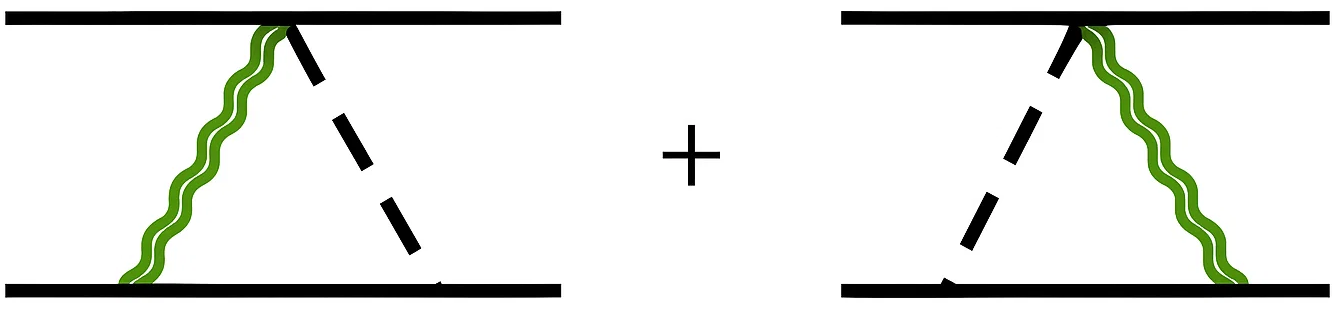}}
    \end{minipage}\hspace{-1 cm}\right] =\int_{\ell}\frac{\mathcal{N}^{(5b)}_{\triangle}}{\rho_1 \rho_2 \rho_3 } \Bigg|_{\textrm{soft}} \nonumber  \\
    & = -\frac{i a^2 m_1 4^{2 \epsilon -3} \pi ^{\epsilon } (\epsilon +1) \left(8 m_2^2+q^2\right) \left(-q^2\right)^{-\epsilon -\frac{1}{2}} \sec (\pi  \epsilon )}{(\epsilon -1) m_p^4 \Gamma (1-\epsilon )}. &
\end{flalign}
\\

\subsection*{Soft amplitudes from penguin topologies  }     
\textbullet $\,\,$ The soft amplitudes for penguin diagrams with two dilaton lines and one graviton line are given by,
\begin{align}
\begin{split}
&\hspace{-0 cm}i\mathbfcal{A}_{4,\textrm{soft}}^{1-\textrm{loop}}\left[\begin{minipage}
          [h]{0.50\linewidth}
	\vspace{0.9 pt}
	\scalebox{0.8}{\includegraphics[width=\linewidth]{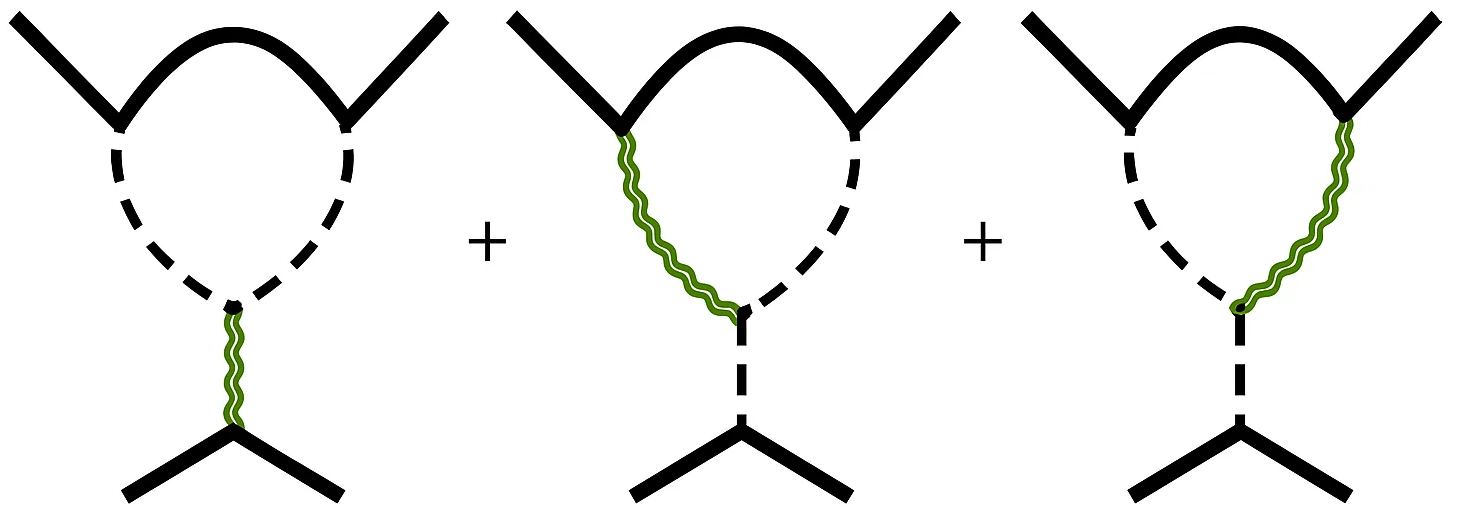}}
    \end{minipage}\hspace{-1.6 cm}\right]=\int_{\ell}\frac{\mathcal{N}_{\penguin}^{(1)}[\ell,q,\bar m_i,\sigma]}{q^2 \rho_1\rho_2\rho_3}\Bigg|_{\textrm{soft}}   \\ &\hspace{-0 cm} =-\frac{i a^2 4^{2 \epsilon -5} \pi ^{\epsilon } \left(-q^2\right)^{-\epsilon -\frac{1}{2}} \sec (\pi  \epsilon )}{m_1 m_p^4 \Gamma (2-\epsilon )}\left[m_2^2 q^2 \left(3 \sigma ^2-1\right)+2 m_1^2 \left(4 m_2^2 \left(\sigma ^2-1\right)+q^2\right)+4 m_1 m_2 q^2 \sigma\right]
      \end{split}
      \end{align}
and,
\begin{align}
\begin{split}
&\hspace{-0 cm}i\mathbfcal{A}_{4,\textrm{soft}}^{1-\textrm{loop}}\left[\begin{minipage}
          [h]{0.50\linewidth}
	\vspace{0.9 pt}
	\scalebox{0.8}{\includegraphics[width=\linewidth]{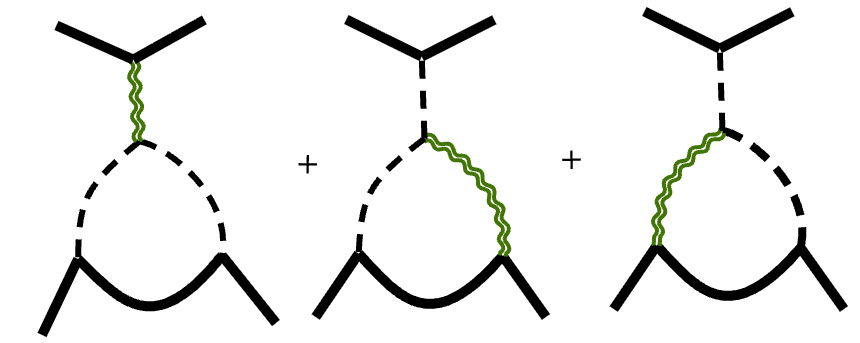}}
    \end{minipage}\hspace{-1.6 cm}\right]=  \int_{\ell}\frac{\mathcal{N}_{\penguin}^{(2)}[\ell,q,\bar m_i,\sigma]}{q^2 \rho_1\rho_2\rho_4}\Bigg|_{\textrm{soft}}  \\ &\hspace{-0 cm} =-\frac{i a^2 4^{2 \epsilon -5} \pi ^{\epsilon } \left(-q^2\right)^{-\epsilon -\frac{3}{2}} \sec (\pi  \epsilon )}{m_2 m_p^4 \Gamma (2-\epsilon )}\left[8 m_1 m_2^2 q^2 \left(m_1 \left(-\sigma ^2+2 \epsilon +1\right)-4 m_2 \epsilon \right)+q^4 \epsilon  \left(8 m_2^2+q^2\right)\right]\,.
      \end{split}
      \end{align}\\
\textbullet $\,\,$ The soft penguin amplitudes for with two photons, one graviton line are given by,
\begin{flalign}
&\hspace{-0 cm}i\mathbfcal{A}_{4,\textrm{soft}}^{1-\textrm{loop}}\left[\begin{minipage}
          [h]{0.50\linewidth}
	\vspace{0.9 pt}
	\scalebox{0.8}{\includegraphics[width=\linewidth]{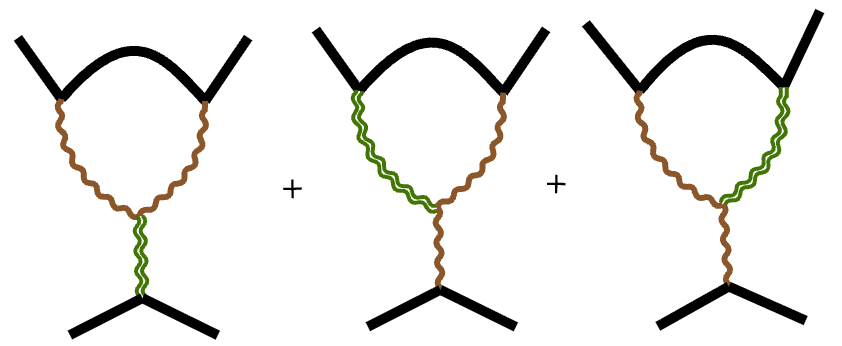}}    \end{minipage}\hspace{-1.6 cm}\right]=\int_{\ell}\frac{\mathcal{N}_{\penguin}^{(3)}[\ell,q,\bar m_i,\sigma]}{q^2 \rho_1\rho_2\rho_3}\Bigg|_{\textrm{soft}}  \nonumber  \\ &\hspace{-0 cm} =-\frac{i \alpha _1 e^2 m_1 m_2 16^{\epsilon -1} \pi ^{\epsilon } \texttt{g}_c \left(-q^2\right)^{-\epsilon -\frac{1}{2}} \sec (\pi  \epsilon ) }{(\epsilon -1) m_p^2 \Gamma (2-\epsilon )} \nonumber \\
    & \hspace{2.5 cm} \times \left[ \alpha _1 m_2 \left(-3 \sigma ^2-4 \sigma ^2 \epsilon ^2+\left(7 \sigma ^2-5\right) \epsilon +1\right)+4 \alpha _2 m_1 \sigma  \left(2 \epsilon ^2-\epsilon +1\right)\right] &
      \end{flalign}
and,
\begin{flalign}
&\hspace{-0 cm}i\mathbfcal{A}_{4,\textrm{soft}}^{1-\textrm{loop}}\left[\begin{minipage}
          [h]{0.50\linewidth}
	\vspace{0.9 pt}
	\scalebox{0.8}{\includegraphics[width=\linewidth]{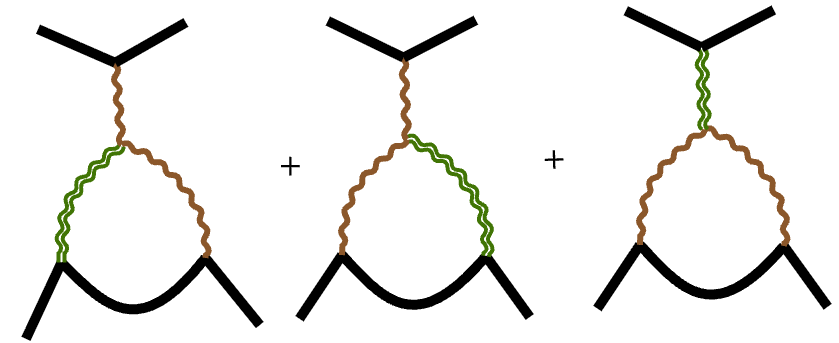}}    \end{minipage}\hspace{-1.6 cm}\right]=\int_{\ell}\frac{\mathcal{N}_{\penguin}^{(4)}[\ell,q,\bar m_i,\sigma]}{q^2 \rho_1\rho_2\rho_4}\Bigg|_{\textrm{soft}} \nonumber   \\ 
    &\hspace{-0 cm} =-\frac{i \alpha _2 e^2 m_1 m_2 16^{\epsilon -1} \pi ^{\epsilon } \texttt{g}_c \left(-q^2\right)^{-\epsilon -\frac{1}{2}} \sec (\pi  \epsilon ) }{(\epsilon -1)^2 m_p^2 \Gamma (1-\epsilon )} \nonumber \\
    & \hspace{2.5 cm} \times \left[ \alpha _2 m_1 \left(3 \sigma ^2+4 \sigma ^2 \epsilon ^2+\left(5-7 \sigma ^2\right) \epsilon -1\right)-4 \alpha _1 m_2 \sigma  \left(2 \epsilon ^2-\epsilon +1\right)\right] \,. &
\end{flalign}\\
\textbullet $\,\,$ The soft amplitudes for the penguin diagram with two photon lines and one dilaton line are given by,
\begin{flalign}
&\hspace{-0 cm}i\mathbfcal{A}_{4,\textrm{soft}}^{1-\textrm{loop}}\left[\begin{minipage}
          [h]{0.50\linewidth}
	\vspace{0.9 pt}
	\scalebox{0.8}{\includegraphics[width=\linewidth]{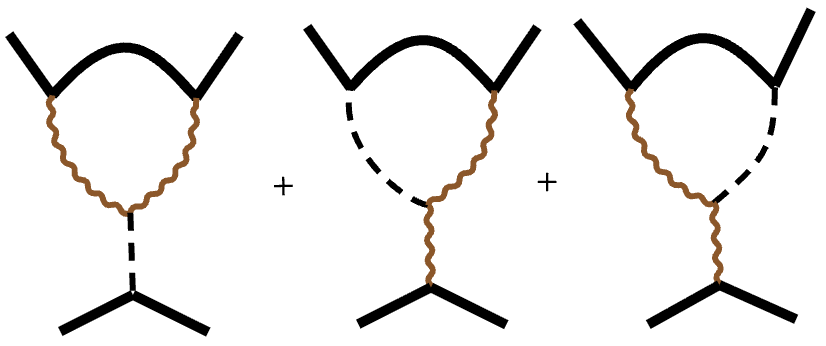}}    \end{minipage}\hspace{-1.6 cm}\right]=\int_{\ell}\frac{\mathcal{N}_{\penguin}^{(5)}[\ell,q,\bar m_i,\sigma]}{q^2 \rho_1\rho_2\rho_3}\Bigg|_{\textrm{soft}}  \nonumber  \\ &\hspace{-0 cm} =\frac{i a \alpha _1 e^2 2^{4 \epsilon -\frac{17}{2}} \pi ^{\epsilon } g_s \left(-q^2\right)^{-\epsilon -\frac{3}{2}} \sec (\pi  \epsilon )}{m_1^3 m_p^2 \Gamma (1-\epsilon )} \nonumber \\ 
    & \hspace{2.5cm} \times \left[128 m_1^4 m_2 q^2 \left(\alpha _1-2 \alpha _2 \sigma \right)-q^4 \left(\alpha _1 m_2+2 \alpha _2 m_1\right) \left(8 m_1^2+q^2\right)\right] &
\end{flalign}
and,
\begin{flalign}
&\hspace{-0 cm}i\mathbfcal{A}_{4,\textrm{soft}}^{1-\textrm{loop}}\left[\begin{minipage}
          [h]{0.50\linewidth}
	\vspace{0.9 pt}
	\scalebox{0.8}{\includegraphics[width=\linewidth]{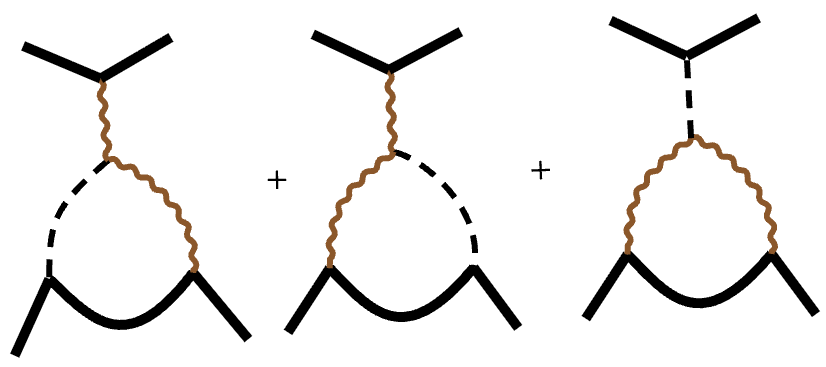}}    \end{minipage}\hspace{-1.6 cm}\right]=\int_{\ell}\frac{\mathcal{N}_{\penguin}^{(6)}[\ell,q,\bar m_i,\sigma]}{q^2 \rho_1\rho_2\rho_4}\Bigg|_{\textrm{soft}} \nonumber   \\ &\hspace{-0 cm} =-\frac{i a \alpha _2 e^2 2^{4 \epsilon -\frac{17}{2}} \pi ^{\epsilon } g_s \left(-q^2\right)^{-\epsilon -\frac{3}{2}} \sec (\pi  \epsilon )}{m_2^3 m_p^2 \Gamma (1-\epsilon )} \nonumber \\
    & \hspace{2.5cm} \times \Big[128 m_1 m_2^4 q^2 \left(2 \alpha _1 \sigma -\alpha _2\right)+q^4 \left(2 \alpha _1 m_2+\alpha _2 m_1\right) \left(8 m_2^2+q^2\right)\Big]\,. &
\end{flalign}\\
\textbullet $\,\,$ The soft amplitudes for penguin topology with three graviton lines are given by,
\begin{flalign}
&\hspace{-0 cm}i\mathbfcal{A}_{4,\textrm{soft}}^{1-\textrm{loop}}\left[\begin{minipage}
          [h]{0.50\linewidth}
	\vspace{1.2 pt}
	\scalebox{0.5}{\includegraphics[width=\linewidth]{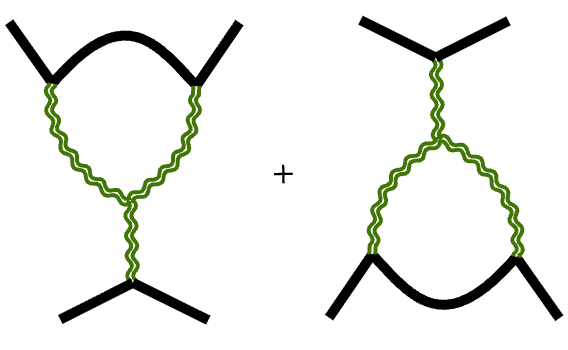}}    \end{minipage}\hspace{-3.7 cm}\right] =\int_{\ell}\frac{\mathcal{N}_{\penguin}^{(7a)}[\ell,q,\bar m_i,\sigma]}{q^2 \rho_1\rho_2\rho_3}\Bigg|_{\textrm{soft}} +\int_{\ell}\frac{\mathcal{N}_{\penguin}^{(7b)}[\ell,q,\bar m_i,\sigma]}{q^2 \rho_1\rho_2 \textcolor{black}{\rho_4}}\Bigg|_{\textrm{soft}}  \nonumber \\ 
    &\hspace{-0 cm} =-\frac{i m_1^2 m_2^2 \left(m_1+m_2\right) 2^{4 \epsilon -9} \pi ^{\epsilon } \left(-q^2\right)^{-\epsilon -\frac{1}{2}}  \sec (\pi  \epsilon )}{(\epsilon -1)^4 m_p^4 \Gamma (1-\epsilon )} \nonumber \\
    & \hspace{3.5cm} \times \left[ -8 \epsilon ^4+11 \epsilon ^3-15 \epsilon ^2+\sigma ^2 (\epsilon -1)^3 (12 \epsilon -1)+\epsilon +3\right] \,. &
\end{flalign}
\noindent   
\hspace{-1.2 cm}The full one loop amplitude in small $|q|$ expansion is given by,\\
\hfsetfillcolor{white!14}
\hfsetbordercolor{black}
\begin{align}
\begin{split}
\tikzmarkin[disable rounded corners=false]{esp1o}(3,-1.7)(-2.4,0.8)
&\hspace{-1.8 cm}i\,\mathbfcal{A}_{\text{1-loop}}(\sigma,|q|,\ep)
= {} (-q^2)^{-(\tfrac32+\ep)}\;\\ & 
\hspace{-1.8 cm}\Biggl[
\;i(-q^2)\;
 \,\frac{2^{-9+4\ep}\,\pi^{\ep}\,\sec(\pi\ep)\,(m_1+m_2)}
             {(-1+\ep)^{3}\,\Gamma(1-\ep)\,\mpow{4}}
\Bigl(
      q^2\bigl(3-9\ep+2\ep^{2}-4\ep^{3}+4(-1+\ep)^{3}\sig-(-1+\ep)^{2}(-1+2\ep)\sig^{2}\bigr)\,m_1 m_2
\\
&\hspace{-1.8 cm}
    +\,2 q^2\bigl(2 \epsilon ^3-\epsilon ^2+4 \epsilon -1\bigr)\,m_2^{2}
    +\,2 m_1^{2}\bigl(q^2(-1+\ep(4+\ep(-1+2\ep))) + 4(-1+\ep)(-\ep+2(-1+\ep)^{2}\sig^{2})\,m_2^{2}\bigr)
\Bigr)
\\&\hspace{-1.8 cm} +\; 
 i\,\frac{16^{-3+\ep}\,\pi^{\ep}\,\sec(\pi\ep)\,(m_1+m_2)}
             {(-1+\ep)^{4}\,\Gamma(1-\ep)\,m_1 m_2\,\mpow{4}}
\Bigl(
   2 q^{6}(-1+\ep)^{3}\ep\,m_2^{2}
\\
&\hspace{-1.8 cm}
 +\, q^{4} m_1 m_2\Bigl( 2 q^{2}
   + q^{2}\bigl(4 \sigma +(6 \sigma +5) \epsilon ^4-22 (\sigma +1) \epsilon ^3+3 (10 \sigma +7) \epsilon ^2-2 (9 \sigma +5) \epsilon\bigr)
   - 8\bigl(\epsilon ^4+4 \epsilon ^3-3 \epsilon ^2+4 \epsilon -2\bigr)m_2^{2}\Bigr)
\\
&\hspace{-1.8 cm}
 +\,2 q^{4} m_1^{2}\Bigl(q^{2}(-1+\ep)^{3}\ep
   + 4\bigl(-2+2\ep+3\ep^{2}-2\ep^{3}+3\ep^{4}+2(-1+\ep)^{3}(-2+3\ep)\sig\bigr)m_2^{2}\Bigr)
\\
&\hspace{-1.8 cm}
 +\,8 m_1^{3} m_2\Bigl(
     -q^{4}\bigl(\epsilon ^4+4 \epsilon ^3-3 \epsilon ^2+4 \epsilon -2\bigr)
     + q^{2}\bigl(3+\ep-15\ep^{2}+11\ep^{3}-8\ep^{4}+(-1+\ep)^{3}(-1+12\ep)\sig^{2}\bigr)m_2^{2}
   \Bigr)
\Bigr)
\\& \hspace{-1.8 cm}+\;
(-q^2)\;
\Bigl\{- i\,\frac{16^{-2+\ep}\,\pi^{\ep}\,\sec(\pi\ep)}
             {(-1+\ep)^{2}\,\Gamma(1-\ep)\,\mpow{4}}
\bigl(
  4 q^{2}(1-\ep)\sig(\ep+2\sig-2\ep\,\sig)\,m_1 m_2 (m_1+m_2)
\\
&\hspace{-1.8 cm}
 -\,(\ep+2\sig-2\ep\,\sig)(m_1+m_2)\bigl(q^{4}(-1+\ep) + 2(1+2(-1+\ep)\sig^{2})\,m_1^{2}m_2^{2}\bigr)
\\
&\hspace{-1.8 cm}
 +\,\frac{q^{2}\bigl(q^{4}(-1+\ep) + 2(1+2(-1+\ep)\sig^{2})\,m_1^{2}m_2^{2}\bigr)}
           {8\,m_1^{2}m_2^{2}}
   \bigl( 2(-1+\ep)\sig\,m_1^{3} + (-8+7\ep)\,m_1^{2} m_2
\\
&\hspace{-1.8 cm}
        +\,(-8+7\ep)\,m_1 m_2^{2} + 2(-1+\ep)\sig\,m_2^{3}\bigr)
\Bigr)\Big\}
\\
&\hspace{-1.8 cm} +\;
(-q^2)^{\tfrac32}\;
\frac{4^{-5+2\ep}\,\pi^{\ep}\,\bigl(1+2(-1+\ep)\sig^{2}\bigr)}
             {(-1+\ep)^{2}\,\mpow{4}}
\Biggl[
  \frac{i\,\sec(\pi\ep)\,(q^{2}+8 m_1^{2})\,m_2^{2}\,(\ep\,m_1+2(-1+\ep)\sig\,m_2)}{\sqrt{-q^2}\,\Gamma(1-\ep)}
\\
&\hspace{-1.8 cm}
 +\,\frac{i\,\sec(\pi\ep)\,m_1^{2}\,(2(-1+\ep)\sig\,m_1+\ep\,m_2)\,(q^{2}+8 m_2^{2})}{\sqrt{-q^2}\,\Gamma(1-\ep)}
\\
&\hspace{-1.8 cm}
 +\,\frac{2\sqrt{\pi}\,\bigl(1+2(-1+\ep)\sig^{2}\bigr)\,(-\sig+\SQ)\,\csc(\pi\ep)\,m_1 m_2\,
         \Bigl(8(\sig^{2}-1)m_1^{2}m_2^{2}-q^{2}(m_1^{2}+2\sig\,m_1 m_2 + m_2^{2})\Bigr)}
        {q^{2}\,(\sig^{2}-1)\,\bigl(1+\sig(-\sig+\SQ)\bigr)\,\Gamma(\tfrac12-\ep)}
\Biggr]
\\ &\hspace{-1.8 cm}\;+\; a^{4}\;
(-q^2)^{\tfrac12}\,
 \frac{2^{-3+4\ep}\,\pi^{\tfrac12+\ep}\,\csc(\pi\ep)\,
          \Bigl(8(\sig^2-1)\,m_1^{2}m_2^{2} - q^2\,(m_1^{2} + 2\sig\,m_1 m_2 + m_2^{2})\Bigr)}
          {(\sig^2-1)^{3/2}\,\Gamma(\tfrac12-\ep)\,m_1 m_2\,\mpow{4}}
\\ &\hspace{-02.5 cm}\;+\; a^{2}\Biggl[
 (-q^2)^{\tfrac32}\;
 \Bigl(-\,\frac{2^{-2+4\ep}\,\pi^{\tfrac12+\ep}\,\sqrt{-q^2}\,\bigl(1-2\sig^2+2\ep\,\sig^2\bigr)\,(-\sig+\SQ)\,\csc(\pi\ep)\,m_1^{2}m_2^{2}}
          {(-1+\ep)\,\bigl(1-\sig^2+\sig\,\SQ\bigr)\,\Gamma(\tfrac12-\ep)\,\mpow{4}}\Bigr)
\\
&\hspace{-1.8 cm}
 +\,(-q^2)\;
 \Bigl(-\, i\,\frac{2^{-4+4\ep}\,\pi^{\ep}\,q^{2}\,(\ep-2\sig+2\ep\,\sig)\,\sec(\pi\ep)\,m_1 m_2\,(m_1+m_2)}
             {(-1+\ep)\,\Gamma(1-\ep)\,\mpow{4}}\Bigr)
\\
&\hspace{-1.8 cm}
 +\,(-q^2)\;\Bigl[
 -\, i\,\frac{4^{-3+2\ep}\,\pi^{\ep}\,(1+\ep)\,\sec(\pi\ep)\,(q^2+8 m_1^{2})\,m_2}
             {(-1+\ep)\,\Gamma(1-\ep)\,\mpow{4}}
 -\, i\,\frac{4^{-3+2\ep}\,\pi^{\ep}\,(1+\ep)\,\sec(\pi\ep)\,m_1\,(q^2+8 m_2^{2})}
             {(-1+\ep)\,\Gamma(1-\ep)\,\mpow{4}}
 \Bigr]
 \tikzmarkend{esp1o}
\end{split}\nonumber
\end{align}
\vspace{-2 cm}
\hfsetfillcolor{white!20}
\hfsetbordercolor{black}
\begin{align}
\begin{split}
\tikzmarkin[disable rounded corners=false]{esp2o}(09,-1.9)(-2.5,1.4)
&\hspace{-1.8 cm}
 +\, i\,\frac{4^{-5+2\ep}\,\pi^{\ep}\,\sec(\pi\ep)\,
          \Bigl(-q^4\,\ep\,(q^2+8 m_1^{2}) + 8 q^2 m_1^{2} m_2\,(4\ep\,m_1 + (-1-2\ep+\sig^2)\,m_2)\Bigr)}
             {\Gamma(2-\ep)\,m_1\,\mpow{4}}
\\
&\hspace{-1.8 cm}
 +\, i\,\frac{4^{-5+2\ep}\,\pi^{\ep}\,\sec(\pi\ep)\,
          \Bigl(8 q^2 m_1 m_2^{2}\bigl((1+2\ep-\sig^2)\,m_1 - 4\ep\,m_2\bigr) + q^4 \ep\,(q^2+8 m_2^{2})\Bigr)}
             {\Gamma(2-\ep)\,m_2\,\mpow{4}}
\\
&\hspace{- 1.8 cm}
 +\,(-q^2)\;
 \Bigl[-\, i\,\frac{2^{-7+4\ep}\,\pi^{\ep}\,\sec(\pi\ep)\,(m_1+m_2)}
             {(-1+\ep)\,\Gamma(1-\ep)\,m_1 m_2\,\mpow{4}}
   \Bigl(q^2(-8+7\ep+2\sig-2\ep\,\sig)\,m_1 m_2
\\
&\hspace{- 1.8 cm}
   +\,2 q^2(-1+\ep)\,\sig\,m_2^{2}
   +\,2 m_1^{2}\bigl(q^2(-1+\ep)\,\sig + (-4\ep-8\sig+8\ep\,\sig)\,m_2^{2}\bigr)\Bigr)\Bigr]
\\
&\hspace{- 1.8 cm}
 +\,(-q^2)\;
 \frac{i\,2^{-5+4\ep}\,\pi^{\ep}\,\sec(\pi\ep)}
             {\Gamma(1-\ep)\,m_1 m_2\,\mpow{4}}
\Bigl[
\begin{aligned}
& q^{2} m_2 (a^{2}-2 b m_2)
 + a^{2} m_1 (q^{2}+8 m_2^{2}) \\
& + m_1^{2}\,\bigl(-2 b q^{2} + 8 m_2(a^{2}-4 b m_2)\bigr)
\end{aligned}
\Bigr]
\Biggr]
\\& \hspace{- 1.8 cm}\;+\; e^{2}\texttt{g}_c\,\alpha_1\alpha_2\Biggl[
 (-q^2)\;
 \Bigl(-\, i\,\frac{2^{-3+4\ep}\,\pi^{\ep}\,\sec(\pi\ep)}
             {(-1+\ep)^{2}\,\Gamma(1-\ep)\,\mpow{2}}\,
   \bigl( q^{2}(1-\ep+\ep^{2})\,m_1 + q^{2}\ep\,\sig\,m_2 + 8(1-\ep+\ep^{2})\,\sig\,m_1^{2} m_2 \bigr)\Bigr)
\\
&\hspace{- 1.8 cm}
 +\,
 \frac{2^{-3+4\ep}\,\pi^{\ep}\,m_1 m_2}
             {(-1+\ep)\,\mpow{2}}
\Biggl[
  -\,\frac{8\sqrt{\pi}\,\sqrt{-q^2}\,\sig\bigl(1+2(-1+\ep)\sig^{2}\bigr)\,\csc(\pi\ep)\,m_1 m_2}{\SQ\,\Gamma(\tfrac12-\ep)}
  \\ &
  \hspace{- 1.8 cm}+\,\frac{i\,q^{2}\bigl(1+2\sig(\ep+3(-1+\ep)\sig)\bigr)\,\sec(\pi\ep)\,(m_1+m_2)}{\Gamma(1-\ep)}
\Biggr]
\\
&\hspace{- 1.8 cm}
 +\,(-q^2)\;
 \frac{i\,2^{-7+4\ep}\,\pi^{\ep}\,\sec(\pi\ep)\,(m_1+m_2)}
             {\Gamma(2-\ep)\,m_1^{3} m_2^{3}\,\mpow{2}}
\Bigl(
 16(1-2\ep\,\sig+6(-1+\ep)\sig^{2})\,m_1^{4} m_2^{4}
\\
&\hspace{- 1.8 cm}
 +\,2 q^{2} m_1^{2} m_2^{2}\Bigl((1+6(-1+\ep)\sig^{2})\,m_1^{2}
    - \bigl(1-6(-8+\sig)\sig + \ep(8-46\sig+6\sig^{2})\bigr)\,m_1 m_2
    + (1+6(-1+\ep)\sig^{2})\,m_2^{2}\Bigr)
\\
&\hspace{- 1.8 cm}
 +\, q^{4}(-1+\ep)\,\Bigl(8 m_1^{2} m_2^{2} + q^{2}(m_1^{2}-m_1 m_2 + m_2^{2})\Bigr)
\Bigr)
\Biggr]
\\&\hspace{- 1.8 cm}\;+\; a^{2} e^{2} \texttt{g}_c\,\alpha_1\alpha_2\Biggl[
 \frac{2^{3+4\ep}\,\pi^{\tfrac12+\ep}\,\sqrt{-q^2}\,\sig(\sig-\SQ)\,\csc(\pi\ep)\,m_1 m_2}
      {\bigl(-1+\sig^{2}-\sig\,\SQ\bigr)\,\Gamma(\tfrac12-\ep)\,\mpow{2}}
\\
&\hspace{- 1.8 cm}
 +\,\frac{i\,2^{4\ep}\,\pi^{\ep}\,q^{2}\,\sec(\pi\ep)\,(m_1+m_2)}{\Gamma(1-\ep)\,\mpow{2}}
 \;+\;
 (-q^2)\;
 \frac{i\,2^{-3+4\ep}\,\pi^{\ep}\,\sec(\pi\ep)\,(m_1+m_2)}
             {\Gamma(1-\ep)\,m_1^{2} m_2^{2}\,\mpow{2}}
\Bigl(-q^{2} m_1 m_2 + q^{2} m_2^{2} + m_1^{2}(q^{2}+8 m_2^{2})\Bigr)
\Biggr]
\\&\hspace{- 1.8 cm}\;+\; e^{4} \texttt{g}_c^{2}\alpha_1^{2}\alpha_2^{2}\Biggl[
 (-q^2)\; i\,\frac{4^{1+2\ep}\,\pi^{\ep}\,\sec(\pi\ep)\,(m_1+m_2)}{\Gamma(1-\ep)}
\\
&\hspace{- 1.8 cm}
 +\,4^{1+2\ep}\,\pi^{\ep}\,\sig
 \Biggl(
    \frac{4\sqrt{\pi}\,\sqrt{-q^2}\,\sig\,(\sig-\SQ)\,\csc(\pi\ep)\,m_1 m_2}{\bigl(1-\sig^{2}+\sig\,\SQ\bigr)\,\Gamma(\tfrac12-\ep)}
   -\frac{i\,q^{2}\,\sec(\pi\ep)\,(m_1+m_2)}{\Gamma(1-\ep)}
 \Biggr)
\\
&\hspace{- 1.8 cm}
 +\,(-q^2)\;
 \Bigl[-\, i\,\frac{2^{-1+4\ep}\,\pi^{\ep}\,\sec(\pi\ep)\,(m_1+m_2)}{\Gamma(1-\ep)\,m_1^{2} m_2^{2}}\,
   \bigl(-q^{2}(-4+\sig)\,m_1 m_2 + q^{2}\sig\,m_2^{2} + \sig\,m_1^{2}(q^{2}+8 m_2^{2})\bigr)\Bigr]
\Biggr]
\\&\hspace{- 1.8 cm}\;+\; e^{2} \texttt{g}_c \Biggl\{
\alpha_1\,
\Bigl[
 \; i\,\frac{4^{-5+2\ep}\,\pi^{\ep}\,\sec(\pi\ep)}
             {\Gamma(2-\ep)\,m_1^{3} m_2^{3}\,\mpow{2}}
 \bigl(
  -q^{2}(-1+3\ep)\,(q^{6}+2 q^{4} m_2^{2})
   \tikzmarkend{esp2o}\label{3.54j}
    \end{split}
\end{align}
\hfsetfillcolor{white!20}
\hfsetbordercolor{black}
\begin{align}
\begin{split}
\tikzmarkin[disable rounded corners=false]{esp3o}(018,-7)(-2.5,1.4)
 \tikzmarkend{esp3o}
 &\hspace{- 1.8 cm}
  + 64 m_1^{4}\bigl(q^{4}\ep(1+\ep) + q^{2}(-1+5\ep + (-1+\ep)(-3+4\ep)\sig^{2})\,m_2^{2}\bigr)
  + 8 m_1^{2}\bigl(q^{6}-3 q^{6}\ep + q^{4}(q^{2}\ep(1+\ep)+(2-6\ep)m_2^{2})\bigr)
 \bigr)
\Bigr]
\\
& \hspace{-1.8 cm} +\;
\alpha_2\,
\Bigl[
 \frac{i\,4^{-5+2\ep}\,\pi^{\ep}\,\sec(\pi\ep)}
      {(-1+\ep)^{2}(\sig-\SQ)^{2}\,\Gamma(1-\ep)\,m_2^{3}\,\mpow{2}}\,
 \bigl(1+2\sig(-\sig+\SQ)\bigr)
\\
&\hspace{- 1.8 cm}
 \times\Bigl(
   (q^{2}+8 m_2^{2})\bigl(q^{6}(1-3\ep)+8 q^{4}\ep(1+\ep)\,m_2^{2}\bigr)\,\alpha_2
 +\,2 m_1^{2}\bigl(32 q^{2}\,m_2^{4} - q^{4}(1-3\ep)(q^{2}+8 m_2^{2})\bigr)\,\alpha_2
\\
&\hspace{-1.8 cm}
 +\,4 \sig\,m_1 m_2\bigl(-64 q^{2}(1+\ep(-1+2\ep))\,m_2^{4}\,\alpha_1
        + q^{4}(-1+\ep)^{2}(q^{2}+8 m_2^{2})(\alpha_1+2\alpha_2)\bigr)
 \Bigr)
\Bigr]
\Biggr\}
\\ &\hspace{-1.8 cm}\;+\; a\,e^{2} \texttt{g}_c \Biggl[
 -\, \frac{i\,2^{-\tfrac{17}{2}+4\ep}\,\pi^{\ep}\,\sec(\pi\ep)}
             {\Gamma(1-\ep)\,m_2^{3}\,\mpow{2}}\,
 \alpha_2 \Bigl( 128 q^{2} m_1\,m_2^{4}\,(2\sig\,\alpha_1 - \alpha_2)
       + q^{4}(q^{2}+8 m_2^{2})\,(2 m_2 \alpha_1 + m_1 \alpha_2) \Bigr)
\\
&\hspace{-1.8 cm}
 +\, \frac{i\,2^{-\tfrac{17}{2}+4\ep}\,\pi^{\ep}\,\sec(\pi\ep)}
             {\Gamma(1-\ep)\,m_1^{3}\,\mpow{2}}\,
 \alpha_1 \Bigl( 128 q^{2} m_1^{4} m_2\,(\alpha_1-2\sig\,\alpha_2)
       - q^{4}(q^{2}+8 m_1^{2})\,(m_2 \alpha_1 + 2 m_1 \alpha_2) \Bigr)
\Biggr]
    +\mathcal{O}(q^2 \log(-q^2))\Biggr]\,.
    \end{split}\nonumber
\end{align}
\vspace{1 cm}
\section{Post-Minkowskian potential: The Lippmann-Schwinger equation and the Infrared subtraction} \label{sec4}   
The Post-Minkowskian potential can be derived from the Lippmann-Schwinger equation. The computation will be easier in the center-of-mass coordinate. The kinematics we choose is the following,
\begin{align}
\begin{split}
&   \texttt{incoming momenta:}\,\, k_1^\mu=(E_1,\boldsymbol{k}),\,k_2^\mu=(E_2,-\boldsymbol{k}),\\ &
\texttt{outgoing momenta:} k_1^{'\mu}=(E_1,\boldsymbol{k+q}), \,k_2^{'\mu}=(E_2,\boldsymbol{-k-q})\,.
\end{split}
\end{align}
One can define the potential in terms of scattering amplitude using the Lippmann-Schwinger equation, which states,
\begin{align}
    \mathbfcal{A}(\boldsymbol{k},\boldsymbol{k'})=V(\boldsymbol{k},\boldsymbol{k'})+\int d^3\boldsymbol{\ell} \,\mathbfcal{A}(\boldsymbol{k},\boldsymbol{\ell}) \,G(\boldsymbol{k},\boldsymbol{\ell})\,V(\boldsymbol{\ell},\boldsymbol{k'}),\label{4.2f}
\end{align}
where $G(\boldsymbol{k},\boldsymbol{\ell})$ is the Green function, and the choice is somewhat arbitrary. However, consistency conditions on scattering amplitude, such as unitarity and demanding that the potential should be a real one, can show (using the optical theorem) that,
\begin{align}
\texttt{Im}\,G(\boldsymbol{k},\boldsymbol{\ell})=\frac{\delta(|\boldsymbol{\ell}|^2-|\boldsymbol{k}|^2)}{16\pi^2 \sqrt{s}}.
\end{align}
The simplest choice of the Green function that satisfies the above is the following 

\begin{align}
\begin{split}
G(\boldsymbol{k},\boldsymbol{\ell})=\frac{1}{(2\pi)^3}\frac{1}{2\sqrt{s}}\frac{1}{|\boldsymbol{\ell}|^2-|\boldsymbol{k}|^2-i0}+\cdots\,.\label{4.4g}
   \end{split}
\end{align}
\textcolor{black}{In principle, one may add to \eqref{4.4g} any term that is regular, provided the constraint is satisfied. Different choices correspond to alternative effective potentials $V$ that differ only by off-shell pieces \cite{Correia:2024jgr}. 
Importantly, the constant piece will lead to a modification of the effective potential in the classical regime. In our computation, we fix the constant piece by comparing it with the traditional Lippmann-Schwinger Green's function, and we will comment on that in detail in the subsequent discussions.}
\begin{figure}
 \hspace{-1 cm}   \centering    \includegraphics[width=0.9\linewidth]{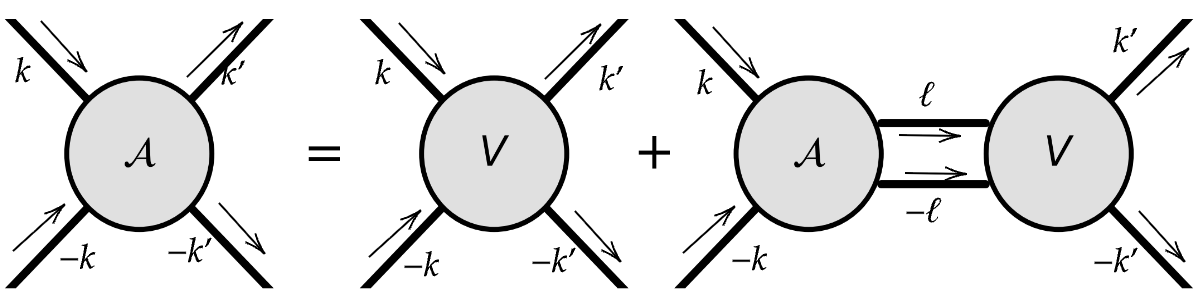}
    \caption{Diagrammatic representation of Lippmann-Schwinger equation}
    \label{fig1}
\end{figure}
Now inverting \eqref{4.2f} and truncate upto one loop order we can write the potential (in momentum space) in terms of scattering amplitude iteratively (see fig. \ref{fig1} for the diagrammatic representation),
\begin{align}
    V(\boldsymbol k,\boldsymbol k')\Big|_{\texttt{1-loop}}=\mathbfcal{A}_{\texttt{1-loop}}(\boldsymbol k,\boldsymbol k')-\int d^3\boldsymbol{\ell}\,\mathbfcal{A}_{\texttt{tree}}(\boldsymbol{k},\boldsymbol{\ell}) G(\boldsymbol{k},\boldsymbol{\ell})\mathbfcal{A}_{\texttt{tree}}(\boldsymbol{\ell},\boldsymbol{k'})\,. \label{4.5h}
\end{align}
Before proceeding to the explicit computation of the potential, let's pause to discuss some intricacies in the choice of the Green's function.
\\
\paragraph{A digression on Green's functions.}
\textcolor{black}{In many amplitude-based derivations (see e.g.~\cite{Cristofoli:2019neg}) one introduces the two-body Green
function
\begin{equation}
  \mathcal{G}(\boldsymbol{k},\boldsymbol{\ell})
  =\frac{1}{E_{\boldsymbol{k}}-E_{\boldsymbol{\ell}}+i0},
  \qquad
  E_{\boldsymbol{\ell}}=\omega_1(\boldsymbol{\ell})+\omega_2(\boldsymbol{\ell})\, .
\end{equation}
In the main text, we instead employ the representation~\eqref{4.4g}, whose leading singular term comes
with the opposite $i0$-prescription. This apparent mismatch is harmless for the conservative observables
we compute: the two choices become equivalent inside the loop integrals once one is allowed to add
terms that are \emph{regular} at $\boldsymbol{\ell}^{2}=\boldsymbol{k}^{2}$ (the ``regular terms'' in~\eqref{4.4g}).
To see this explicitly, Taylor-expand $\mathcal{G}(\boldsymbol{k},\boldsymbol{\ell})$ around
$\boldsymbol{\ell}\simeq\boldsymbol{k}$. Writing the energy difference as an expansion in
$\Delta\equiv(\boldsymbol{\ell}^{2}-\boldsymbol{k}^{2})$, one finds}
\textcolor{black}{\begin{align}
  E_{\boldsymbol{k}}-E_{\boldsymbol{\ell}}+i0
  &=
  -\left.\frac{\partial E_{\boldsymbol{\ell}}}{\partial \boldsymbol{\ell}^{2}}\right|_{\boldsymbol{\ell}^{2}=\boldsymbol{k}^{2}} \Delta
  -\frac12\left.\frac{\partial}{\partial \boldsymbol{\ell}^{2}}
  \Big(\frac{\partial E_{\boldsymbol{\ell}}}{\partial \boldsymbol{\ell}^{2}}\Big)\right|_{\boldsymbol{\ell}^{2}=\boldsymbol{k}^{2}}\Delta^{2}
  +\cdots + i0\,, \nonumber\\[2pt]
  \left.\frac{\partial E_{\boldsymbol{\ell}}}{\partial \boldsymbol{\ell}^{2}}\right|_{\boldsymbol{\ell}^{2}=\boldsymbol{k}^{2}}
  &=\frac{1}{2\,\xi(\boldsymbol{k})\,E_{\boldsymbol{k}}}\,,\qquad
  \left.\frac{\partial}{\partial \boldsymbol{\ell}^{2}}
  \Big(\frac{\partial E_{\boldsymbol{\ell}}}{\partial \boldsymbol{\ell}^{2}}\Big)\right|_{\boldsymbol{\ell}^{2}=\boldsymbol{k}^{2}}
  =-\frac{1-3\xi(\boldsymbol{k})}{4\,\xi(\boldsymbol{k})^{3}\,E_{\boldsymbol{k}}^{3}}\, .
\end{align}}
\textcolor{black}{
In the large center-of-mass energy regime (large $E_{\boldsymbol{k}}$) this yields
\begin{equation}
  \mathcal{G}(\boldsymbol{k},\boldsymbol{\ell})
  =-\frac{2E_{\boldsymbol{k}}\,\xi(\boldsymbol{k})}{\boldsymbol{\ell}^{2}-\boldsymbol{k}^{2}+i0}
  +\frac{3\xi(\boldsymbol{k})-1}{2\,\xi(\boldsymbol{k})\,E_{\boldsymbol{k}}}
  +\mathcal{O}\!\left(\frac{\boldsymbol{\ell}^{2}-\boldsymbol{k}^{2}}{E_{\boldsymbol{k}}^{3}}\right),
  \qquad
  \xi(\boldsymbol{k})=\frac{\omega_1(\boldsymbol{k})\omega_2(\boldsymbol{k})}{(\omega_1(\boldsymbol{k})+\omega_2(\boldsymbol{k}))^{2}}\, .
  \label{eq:Gexp-large-s}
\end{equation}
On the other hand, our choice~\eqref{4.4g} has the leading behavior}
\textcolor{black}{
\begin{equation}
  G(\boldsymbol{k},\boldsymbol{\ell})
  =\frac{1}{(2\pi)^{3}}\frac{1}{2\sqrt{s}}\,
  \frac{1}{\boldsymbol{\ell}^{2}-\boldsymbol{k}^{2}-i0}+\cdots
  \;\;\longrightarrow\;\;
  \frac{1}{(2\pi)^{3}}\frac{1}{2E_{\boldsymbol{k}}}\,
  \frac{1}{\boldsymbol{\ell}^{2}-\boldsymbol{k}^{2}-i0}+\cdots\, ,
  \label{eq:G-ours-leading}
\end{equation}
which differs from~\eqref{eq:Gexp-large-s} (up to normalization) by an overall sign and by the sign of the
$i0$-prescription. Nevertheless, both prescriptions lead to the same integrated result for the class of
amplitudes we need, namely
\begin{equation}
  \mathcal{A}=\int\!\frac{d^{d}\boldsymbol{\ell}}{(2\pi)^{d}}\,
  \frac{\mathrm{GF}(\boldsymbol{k},\boldsymbol{\ell})}{|\boldsymbol{\ell}-\boldsymbol{k}|^{2}\,|\boldsymbol{\ell}-\boldsymbol{k}'|^{2}}\, .
\end{equation}
Indeed, after the standard manipulations, one encounters an integral of the schematic form
\begin{equation}
  \mathcal{A}[G]=
  \int d^{d}\boldsymbol{\ell}\,
  \frac{1}{\big(\boldsymbol{\ell}\!\cdot\!\tilde{\boldsymbol{q}}_{\perp}-i0\big)\,\boldsymbol{\ell}^{2}\,(\boldsymbol{\ell}-\boldsymbol{q})^{2}}\, .
\end{equation}
Now perform the change of variables $\boldsymbol{\ell}\to-\boldsymbol{\ell}$ together with
$\boldsymbol{q}\to-\boldsymbol{q}$ (which leaves the integration domain invariant). This gives
\begin{equation}
  \mathcal{A}[G]=
  -\int d^{d}\boldsymbol{\ell}\,
  \frac{1}{\big(\boldsymbol{\ell}\!\cdot\!\tilde{\boldsymbol{q}}_{\perp}+i0\big)\,\boldsymbol{\ell}^{2}\,(\boldsymbol{\ell}-\boldsymbol{q})^{2}}
  \equiv \mathcal{A}[\mathcal{G}]\, ,
\end{equation}
showing that the two seemingly different Green functions yield the same result once integrated.
Finally, the ellipsis in~\eqref{eq:G-ours-leading} can be fixed by matching to~\eqref{eq:Gexp-large-s} up to
terms analytic at $\boldsymbol{\ell}^{2}=\boldsymbol{k}^{2}$; a convenient choice is
\begin{equation}
  G(\boldsymbol{k},\boldsymbol{\ell})=\frac{1}{(2\pi)^{3}}
  \left[
    \frac{1}{2E_{\boldsymbol{k}}(\boldsymbol{\ell}^{2}-\boldsymbol{k}^{2}-i0)}
    +\frac{3\xi(\boldsymbol{k})-1}{8\,\xi^2(\boldsymbol{k})\,E_{\boldsymbol{k}}^3}
    +\mathcal{O}\!\left(\frac{\boldsymbol{\ell}^{2}-\boldsymbol{k}^{2}}{E_{\boldsymbol{k}}^{5}}\right)
  \right],
\end{equation}
where the subleading pieces are precisely the ``regular terms'' alluded to below~\eqref{4.4g}.}
\\\\
Therefore, the potential in position space can be computed by taking the Fourier transform of \eqref{4.5h}. We have the one-loop and tree-level amplitude. Then our first task to compute the iterative amplitude  \eqref{4.5h}, which yields (we now work in $d$ dimension to isolate the IR divergence),
\begin{align}
  \mathbfcal{A}_{\texttt{Iterative}}=  \int d^d\boldsymbol{\ell}\,\mathbfcal{A}_{\texttt{tree}}(\boldsymbol{k},\boldsymbol{\ell}) G(\boldsymbol{k},\boldsymbol{\ell})\mathbfcal{A}_{\texttt{tree}}(\boldsymbol{\ell},\boldsymbol{k'})
\end{align}
where the tree amplitude is given by,
\begin{align}
\begin{split}
    \mathbfcal{A}_{\texttt{tree}}(\boldsymbol{k},\boldsymbol{\ell})=&-\frac{1}{|\boldsymbol{\ell}-\boldsymbol{k}|^2}\left(-\frac{4  a^2 m_1 m_2}{m_p^2}+16  \alpha _1 \alpha _2 \texttt{g}_ce^2 m_1 m_2 \sigma  +\frac{ m_1^2 m_2^2 \left(-2 \sigma ^2 +1\right)}{2 m_p^2}\right)\,,\\ &
    +\left(4  \alpha _1 \alpha _2 \texttt{g}_c e^2 -\frac{ 2 m_2 m_1 \sigma   }{4  m_p^2}\right)\,,\\ &
    =-c_{1}(\sigma)\frac{1}{|\boldsymbol{\ell}-\boldsymbol{k}|^2}+c_2(\sigma)\,.
    \end{split}
    \end{align}
Therefore, the iterative Born-subtracted amplitude is given by,
\begin{align}
    \begin{split}
        \mathbfcal{A}_{\texttt{Iterative}}=\mathbfcal{A}^{(1)}_{\texttt{Iterative}}+\mathbfcal{A}^{(2)}_{\texttt{Iterative}}+\mathbfcal{A}^{(3)}_{\texttt{Iterative}}
    \end{split}
\end{align}
where,
\begin{align}
    \begin{split}\mathbfcal{A}^{(1)}_{\texttt{Iterative}}(\boldsymbol{k},\boldsymbol{k'})&= \frac{c_1^2(\sigma)}{2E_{\boldsymbol{k}}} \int \frac{d^d\boldsymbol{\ell}}{(2\pi)^3}\,\frac{1}{|\boldsymbol{\ell}-\boldsymbol{k}|^2(\boldsymbol{\ell}^2-\boldsymbol{k}^2-i0)|\boldsymbol{\ell}-\boldsymbol{k'}|^2}\,,\\ &
    =\frac{c_1^2(\sigma)}{2E_{\boldsymbol{k}}}  \int \frac{d^d\boldsymbol{\ell}}{(2\pi)^3}\frac{1}{(\boldsymbol{\ell}^2+2\boldsymbol{\ell}\cdot \boldsymbol{k}-i0)\boldsymbol{\ell}^2(\boldsymbol{\ell}-\boldsymbol{q})^2}\,.\label{4.20jj}
    \end{split}
\end{align}
As before, the integral can be done using the soft loop expansion.
\begin{align}
    \begin{split}\int_{\boldsymbol{\ell\sim\boldsymbol{q}}} \frac{d^d\boldsymbol{\ell}}{(2\pi)^3}\frac{1}{(\boldsymbol{\ell}^2+2\boldsymbol{\ell}\cdot \boldsymbol{k}-i0)\boldsymbol{\ell}^2(\boldsymbol{\ell}-\boldsymbol{q})^2}&=\int_{\boldsymbol{\ell\sim\boldsymbol{q}}} \frac{d^d\boldsymbol{\ell}}{(2\pi)^3}\frac{1}{(\boldsymbol{\ell}^2+2\boldsymbol{\ell}\cdot (\frac{\tilde{\boldsymbol{q}}_{\perp}-\boldsymbol{q}}{2})-i0)\boldsymbol{\ell}^2(\boldsymbol{\ell}-\boldsymbol{q})^2}\,,\\ &    \hspace{-2.5 cm}=\int_{\boldsymbol{\ell\sim\boldsymbol{q}}} \frac{d^d\boldsymbol{\ell}}{(2\pi)^3}\left(1-\frac{\boldsymbol{\ell}^2-\boldsymbol{\ell}\cdot \boldsymbol{q}}{\boldsymbol{\ell}\cdot\tilde{\boldsymbol{q}}_{\perp}}+\cdots\right) \frac{1}{(\boldsymbol{\ell}\cdot\boldsymbol{q_{\perp}}-i0)\boldsymbol{\ell}^2(\boldsymbol{\ell-q})^2}\label{4.10o}
    \end{split}
\end{align}
where, $\tilde{\boldsymbol{q}}_{\perp}=\boldsymbol{q}+2\boldsymbol{k}$ and also note that $\tilde{\boldsymbol{q}}_{\perp}\cdot \boldsymbol{q}=0$. Keeping this in mind the first term in \eqref{4.10o} takes the form,
\begin{align}
    \textstyle{\mathbfcal{A}^{(1)}_{\texttt{Iterative}}(\boldsymbol{k},\boldsymbol{k'})=\frac{c^2_1(\sigma)}{2E_{\boldsymbol{k}}}\int_{\boldsymbol{\ell\sim\boldsymbol{q}}} \frac{d^d\boldsymbol{\ell}}{(2\pi)^3} \frac{1}{(\boldsymbol{\ell}\cdot\tilde{\boldsymbol{q}}_{\perp}-i0)\boldsymbol{\ell}^2(\boldsymbol{\ell-q})^2}}&\xrightarrow[\boldsymbol{q}\to -\boldsymbol{q}]{\boldsymbol{\ell}\to -\boldsymbol{\ell}}-\frac{c^2_{1}(\sigma)}{2E_k}   \int d^{d}\boldsymbol{\ell}\frac{1}{(\boldsymbol{\ell}\cdot \tilde{\boldsymbol{q}}_{\perp}+i0)\boldsymbol{\ell}^2(\boldsymbol{\ell}-\boldsymbol{q})^2},
\end{align}
and adding the two, we get,
\begin{align}
   \begin{split}\mathbfcal{A}^{(1)}_{\texttt{Iterative}}(\boldsymbol{k},\boldsymbol{k'})= \frac{\pi i c_1^2(\sigma)}{2E_{\boldsymbol{k}}}\int \frac{d^d\boldsymbol{\ell}}{(2\pi)^3} \frac{\delta(\boldsymbol{\ell}\cdot\tilde{\boldsymbol{q}}_{\perp})}{\boldsymbol{\ell}^2(\boldsymbol{\ell}-\boldsymbol{q})^2}&=\frac{c_1^2(\sigma)}{2E_{\boldsymbol{k}}}\frac{i\pi}{2\pi |\tilde{\boldsymbol{q}}_{\perp}|}\int \frac{d^{d-1}}{(2\pi)^2}\boldsymbol{\ell} \frac{1}{\boldsymbol{\ell}^2(\boldsymbol{\ell}-\boldsymbol{q})^2}\,,\\ &
 =\frac{c_1^2(\sigma)}{2E_{\boldsymbol{k}}}\frac{i}{2|\boldsymbol{q}_\perp|}\left(\frac{ \log (|\boldsymbol{q}|^2)+\gamma_{E} -\log (4 \pi )}{2 \pi  |\boldsymbol{q}|^2}-\frac{1}{2 \pi |\boldsymbol{ q}|^2 \epsilon }\right)\,.
    \end{split}
\end{align}
Again note that, in the soft limit $|\boldsymbol{q}_\perp|=\sqrt{4\boldsymbol{k}^2-\boldsymbol{q}^2}\sim {2}|\boldsymbol{k}|$.
As we immediately see that the iterative amplitude has an IR divergent piece which is given by,
\begin{align}
    \mathbfcal{A}_{\texttt{Iterative}}^{(1)}(\boldsymbol{k},\boldsymbol{k'})\Bigg|_{\texttt{IR div.}}= -\frac{i c_1^2(\sigma)}{16 \pi E_{\boldsymbol{k}}|\boldsymbol{k}||\boldsymbol{q}|^2\,\epsilon}=-\frac{ic_1^2(\sigma)}{16\pi m_1 m_2 \sqrt{\sigma^2-1}|\boldsymbol{q}|^2\epsilon}\,.
\end{align}
On the other hand,  the one-loop amplitude computed in \eqref{3.54j} has an IR divergent piece,
\begin{align}
    \begin{split}\mathbfcal{A}_{\texttt{1-loop}}\Bigg|_{\texttt{IR div.}}&=-\frac{i m_1 m_2 \left(8 \left(a^2-4 \alpha _1 \alpha _2 e^2 \sigma  \texttt{g}_c m_p^2\right)+m_1 m_2 \left(2 \sigma ^2-1\right)\right){}^2}{64 \pi  \sqrt{\sigma ^2-1} \epsilon \, m_p^4 |\boldsymbol{q}|^2}\,,\\ & =\frac{-ic_1^2(\sigma)}{16\pi m_1 m_2\sqrt{\sigma^2-1}|\boldsymbol{q}|^2\epsilon}\to \mathbfcal{A}_{\texttt{Iterative}}^{(1)}(\boldsymbol{k},\boldsymbol{k'})\Bigg|_{\texttt{IR div.}} \,.
    \end{split}
\end{align}
From the above analysis, \textit{We observe a complete cancellation of infrared singularities: the IR divergence present in the one-loop amplitude is exactly removed by the iterative (Born) contribution. Hence, the Born (IR)-subtracted Post-Minkowskian potential is IR finite. \textcolor{black}{The same subtraction scheme has been first applied for general relativity in~\cite{Cristofoli:2019neg}, where the Born-subtracted potential was also demonstrated to be IR finite.}}   As we will see, the other integrals appearing in the iterative amplitude are free of IR divergence.
Similarly, we have,
\begin{align}
    \begin{split}
        \mathbfcal{A}_{\texttt{Iterative}}^{(2)}(\boldsymbol{k},\boldsymbol{k'})&\sim-c_1(\sigma)c_2(\sigma)\int d^d\boldsymbol{\ell}\frac{1}{|\boldsymbol{\ell}-\boldsymbol{k}|^2(\boldsymbol{\ell}^2-\boldsymbol{k}^2)}\,,\\ &
        =-c_1(\sigma)c_2(\sigma)\int d^d\boldsymbol{\ell}\frac{1}{\boldsymbol{\ell}^2(\boldsymbol{\ell}^2+2\boldsymbol{\ell}\cdot\boldsymbol{k}-i0)}=0\,\, (\texttt{scaleless in soft limit})\,,\\ &
        \to \mathbfcal{A}_{\texttt{Iterative}}^{(3)}(\boldsymbol{k},\boldsymbol{k'}) \,\,(\texttt{in soft limit})\,.
    \end{split}
\end{align}
However, we are not completely done yet. We see that the iterative Born subtraction only has a superclassical contribution. However, we still have the freedom to add a regular term coming from the Taylor expansion mentioned in the footnote. The Green function takes the form,
\begin{align}
    G(\boldsymbol{k},\boldsymbol{\ell})=\frac{1}{(2\pi)^3}\frac{1}{2\sqrt{s}}\frac{1}{|\boldsymbol{\ell}|^2-|\boldsymbol{k}|^2-i0}+\frac{3\xi-1}{64 \pi^3 \xi^2 {s^{\frac{3}{2}}}}+\cdots\,.
\end{align}
Therefore, the only surviving correction to the iterative amplitude in the soft limit is the following,
\begin{align}
    \begin{split}
        \mathbfcal{A}_{\texttt{Iterative}}'(\boldsymbol{k},\boldsymbol{k'})=c_1^2(\sigma)\frac{3\xi-1}{64\pi^3\xi^2s^{\frac{3}{2}}}\int \frac{d^d\boldsymbol{\ell}}{\boldsymbol{\ell}^2(\boldsymbol{\ell}-\boldsymbol{q})^2}=c_1^2(\sigma)\frac{3\xi-1}{64\xi^2s^{\frac{3}{2}}|\boldsymbol{q}|}\,.
    \end{split}
\end{align}
Therefore, the one-loop classical potential in momentum space is given by,
\begin{align}
V(\boldsymbol{k},|\boldsymbol{q}|)=\mathbfcal{A}_{\texttt{1-loop}}(\boldsymbol{k},|\boldsymbol{q}|)\Big|_{\frac{1}{|\boldsymbol{q}|}}-\frac{c_1^2(\sigma)(3\xi-1)}{64\xi^2 E_{\boldsymbol{k}}^3 |\boldsymbol{q}| }\,.
\end{align}
The two-body classical  potential can be computed by taking the Fourier transform of the momentum space amplitude,
\begin{align}   
\begin{split}
V_{\texttt{1-loop}}(\boldsymbol{k},|\boldsymbol{r}|)=-\frac{1}{
4m_1m_2}\int \frac{d^3\boldsymbol{q}}{(2\pi)^3}e^{i\boldsymbol{q}\cdot \boldsymbol{r}}\,V(\boldsymbol{k},|\boldsymbol{q}|) =-\frac{\mathcal{C}_2(\sigma)}{8m_1m_2\pi^2 r^2}\,.
\end{split}
\end{align}
where,
\begin{align}
    \begin{split}
         \mathcal{C}_{2}(\sigma) &= \frac{a^4 \left(m_1+m_2\right)}{4 m_p^4}-\frac{a^2 m_1 m_2 \left(128 b+\left(m_1+m_2\right) \left(\sigma ^2-17\right)\right)}{128 m_p^4}\\
         & -\frac{\sqrt{2} a e^2 m_1 m_2 \left(-4 \alpha _2 \alpha _1 \sigma +\alpha _1^2+\alpha _2^2\right) \texttt{g}_c}{4 m_p^2}  +\frac{3 m_1^2 m_2^2 \left(m_1+m_2\right) \left(5 \sigma ^2-1\right)}{512 m_p^4} \\
         &+ 4 \alpha _1^2 \alpha _2^2 e^4 \left(m_1+m_2\right) \texttt{g}_c^2 -\frac{e^2 m_1 m_2 \texttt{g}_c \left(\left(3 \sigma ^2-1\right) \left(\alpha _1^2 m_2+\alpha _2^2 m_1\right)+12 \alpha _1 \alpha _2 \left(m_1+m_2\right) \sigma \right)}{16 m_p^2}\\
         & -\frac{c_1^2(\sigma)(3\xi-1)}{64\xi^2 E_{\boldsymbol{k}}^3  } \,,\label{4.21g}
    \end{split}
\end{align}
with{\footnote{where $a,b$ has mass dimension 1.}} $\sigma=\frac{k_1\cdot k_2}{m_1m_2}\,$. If one turns off the dilaton field and takes the static limit, then the potential written in \eqref{4.21g} agrees with \cite{Bjerrum-Bohr:2002aqa,Faller:2007sy}. Before closing this section, we briefly comment on a mild ambiguity that arises from the freedom to add \emph{regular} (analytic) terms to the Green's function, as mentioned earlier.\\\\
\textcolor{black}{
\textbf{Comment on adding regular pieces in the Green's function.}  The freedom to add \emph{regular} (analytic) terms to the Green’s function only affects the off-shell extension of the integrand and can at most generate
shorter-range contributions to the conservative potential. In particular, such terms do \emph{not} modify the IR-singular piece that is removed by the Born/iteration subtraction.
To see this explicitly, note that adding analytic regular terms produces (schematically) loop integrals of the form
\begin{align}
\mathcal{I}_n(\boldsymbol{q},\boldsymbol{q}_\perp)
\;\equiv\;
\int d^d\boldsymbol{\ell}\;
\frac{\bigl(2\,\boldsymbol{\ell}\!\cdot\!\boldsymbol{q}_\perp\bigr)^n}{\boldsymbol{\ell}^{\,2}\,(\boldsymbol{\ell}-\boldsymbol{q})^{2}}\,,
\qquad n\ge 1,
\label{eq:In_def_refreply}
\end{align}
where $\boldsymbol{q}_\perp$ is transverse, $\boldsymbol{q}\!\cdot\!\boldsymbol{q}_\perp=0$ (mass on-shell condition: $|\boldsymbol{k}|^2=|\boldsymbol{k}+\boldsymbol{q}|^2$). By symmetry, $\mathcal{I}_n$ vanishes identically for odd $n$.
For even $n$, a standard IBP reduction implies that the numerator can only generate powers of
$\boldsymbol{q}^2$ and $\boldsymbol{q}_\perp^{\,2}$, so that one may write
\begin{align}
\mathcal{I}_n(\boldsymbol{q},\boldsymbol{q}_\perp)
=
C_n(d)\,(\boldsymbol{q}^{2})^{n/2}(\boldsymbol{q}_\perp^{\,2})^{n/2}
\int d^d\boldsymbol{\ell}\,\frac{1}{\boldsymbol{\ell}^{\,2}(\boldsymbol{\ell}-\boldsymbol{q})^{2}}
\;\propto\;
(\boldsymbol{q}^2)^{\frac{n-1}{2}}
\qquad (\text{even } n),
\label{eq:In_scaling_refreply}
\end{align}
where in the last step we used the well-known scaling of the massless bubble integral in $d=3-2\epsilon$ spatial dimensions,
$\int d^d\boldsymbol{\ell}\, [\boldsymbol{\ell}^2(\boldsymbol{\ell}-\boldsymbol{q})^2]^{-1}\propto (\boldsymbol{q}^2)^{d/2-2}\sim (\boldsymbol{q}^2)^{-1/2-\epsilon}$.
Crucially, for $n\ge 1$ the behavior \eqref{eq:In_scaling_refreply} is \emph{less singular} in the soft limit than the $n=0$ term, and therefore it does not contribute to the IR divergence that is cancelled by the iteration (Born) subtraction. Finally, the corresponding contribution to the conservative potential is obtained by Fourier transform.
Using $\int d^3\boldsymbol{q}\,e^{i\boldsymbol{q}\cdot\boldsymbol{r}}\,|\boldsymbol{q}|^{\,m}\sim |r|^{-m-3}$ (up to constants), one finds
\begin{align}
V_n(r)\;\sim\;\int d^3\boldsymbol{q}\;e^{i\boldsymbol{q}\cdot\boldsymbol{r}}\,
(\boldsymbol{q}^2)^{\frac{n-1}{2}}
\;\propto\;\frac{1}{|\boldsymbol{r}|^{\,n+2}}\,, \qquad n=2,4,6,\cdots,
\label{eq:Vn_falloff_refreply}
\end{align}
i.e.\ a contribution that is parametrically shorter-ranged (and more suppressed at large $r$) than the leading long-range terms.
Therefore, purely analytic regular terms cannot alter the IR-subtracted \emph{long-range} potential: their effect is confined to short-distance/contact contributions and, more generally, to off-shell terms that can be absorbed by local field redefinitions (or equivalently by canonical transformations).
It is useful to recall how the relevant non-analytic structures in the momentum-transfer expansion map into the large-distance behavior of the potential at $\mathcal{O}(G^2)$.
In three spatial dimensions, the Fourier transform implies the schematic correspondences
\begin{align}
\frac{1}{|\boldsymbol{q}|^{2}}\;\longleftrightarrow\;\frac{G^2}{r\hbar}\,,\qquad
\frac{1}{|\boldsymbol{q}|}\;\longleftrightarrow\;\frac{G^2 \hbar^0}{r^{2}}\,,\qquad
\log(\boldsymbol{q}^{2})\;\longleftrightarrow\;\frac{G^2\hbar}{r^{3}}\,,
\end{align}
which are naturally interpreted as the super-classical, classical, and quantum contributions, respectively.
In our analysis, the apparent super-classical ($\sim 1/\hbar$) pole is spurious and is removed by the standard leading Born subtraction, leaving an IR-finite long-range result.
Regular terms, on the other hand, only introduce an ambiguity in the definition of the post-Minkowskian potential at the level of off-shell completion.
As emphasized in \cite{Correia:2024jgr,Correia:2024yfx}, different choices of regular terms lead to potentials that differ by off-shell pieces, and hence they do not affect on-shell, gauge-invariant observables (such as the scattering angle or other conservative observables derived from the S-matrix).  }
Next, we discuss eikonal exponentiation in momentum space and deduce the scattering angle.

\section{Eikonal exponentiation and the scattering angle}\label{sec5}
In this section, we will study the eikonal exponentiation of the conservative scattering amplitude directly in the momentum space following \cite{Parra-Martinez:2020dzs}. Traditionally, in eikonal exponentiation, it is well known that the eikonal phase can be computed by taking the Fourier transform of the scattering amplitude in impact-parameter space.
The primary reason to do the computation in momentum space is to avoid additional IR divergence coming from the tree-level amplitudes arising from the Coulomb interaction. However, one must pay the computational cost in momentum space: simple products in impact parameter space give rise to convolutions in momentum space, which can be computed using the iterative bubble integrals. The statement of eikonal exponentiation reduces to the following: one can write the scattering amplitude as a \textit{convolutional exponential} of the eikonal phase,
\begin{align}
\begin{split}
    i\mathbfcal{A}(\sigma,-q^2)&=\textrm{cexp}\left(i\delta(\sigma,-q^2)\right)-1\,,\\ &
    =i\delta(\sigma,-q^2)-\frac{1}{2!}\delta(\sigma,-q^2)\otimes \delta(\sigma,-q^2)-i\frac{1}{3!}\delta(\sigma,-q^2)\otimes \delta(\sigma,-q^2)\otimes\delta(\sigma,-q^2)+\cdots
    \end{split}
\end{align}
where the convolution is defined as an integral over the $d-2$ dimensional transverse space.
\begin{align}    \varpi_1(\boldsymbol{q}_{\perp})\otimes \varpi_2(\boldsymbol{q}_{\perp}):=\frac{1}{N}\int d\mu_{\boldsymbol{\ell}_{\perp}}^{d-2}\,\varpi_1(\boldsymbol{q}_{\perp})\varpi_2(\boldsymbol{q}_{\perp}-\boldsymbol{\ell}_{\perp})\,,
\end{align}
where $N=4 m_1 m_2 \sqrt{\sigma^2-1}$ is the normalization factor. Similarly, one can write the inverse relation as,
\begin{align}
    \begin{split}
        \delta(\sigma,-q^2)&=-i\textrm{clog}(1+i \mathbfcal{A})\,, \\
        &=\mathbfcal{A}(\sigma ,-q^2)-\frac{i}{2}\mathbfcal{A}(\sigma ,-q^2) \otimes\mathbfcal{A}(\sigma ,-q^2) -\frac{1}{3}\mathbfcal{A}(\sigma ,-q^2) \otimes \mathbfcal{A}(\sigma ,-q^2) \otimes\mathbfcal{A}(\sigma ,-q^2)+\cdots\,.
    \end{split}
\end{align}
This can be written as 
\begin{align}
    \begin{split}
        \delta= \delta^{(0)}+\delta^{(1)}+\delta^{(2)}+\cdots
    \end{split}
\end{align}
where $ \delta^{(L)}$ has $\mathcal{O}(e^{2L+2})$,  $\mathcal{O}(G^{L+1})$ and $\mathcal{O}(e^{2L} G^{L})$ terms and so on hence we can write phases as
\begin{align}
    \begin{split}
    &\delta^{(0)}=\mathbfcal{A}^{\textrm{tree}}\,, \\
    & \delta^{(1)}=\mathbfcal{A}^{\textrm{1-loop}} -\frac{i}{2} \mathbfcal{A}^{\textrm{tree}} \otimes \mathbfcal{A}^{\textrm{tree}}\,.
    \end{split}
\end{align}
So we have
\begin{align}
    \delta^{(0)}(\sigma, -q^2)= \frac{1}{q^2} \left( -\frac{4 a^2 m_1 m_2}{m_p^2} + 16 \alpha _1 \alpha _2 e^2   m_1 m_2 \sigma   \texttt{g}_c+\frac{m_1^2 m_2^2 \left(-2 \sigma ^2+2 \sigma ^2 \epsilon +1\right)}{(2-2 \epsilon ) m_p^2} \right) \label{del0p}
\end{align}
where the $\mathcal{O}(-q^0)$ term of $\mathbfcal{A^{\textrm{tree}}}$ won't contribute as they lead to $\delta(\boldsymbol{b_e})$ (which is simply zero as $\boldsymbol{b_e} \neq 0$) when fourier transformed to impact parameter space.
Now using 
\begin{align}
    \frac{1}{-q^2}\otimes\frac{1}{-q^2}=\frac{1}{\boldsymbol{q^2_\perp}}\otimes\frac{1}{\boldsymbol{q^2_\perp}}=\frac{1}{N} (4 \pi )^{\epsilon -1}  \frac{  \Gamma (-\epsilon )^2 \Gamma (\epsilon +1)}{\Gamma (-2 \epsilon )} (\boldsymbol{q}^2_\perp)^{-\epsilon -1}
\end{align}
we get the following, 
\begin{align}
    \begin{split}
        &\delta^{(1)}(\sigma, -q^2) = \frac{i \left(2 m_2 m_1 \sigma +m_1^2+m_2^2\right) \left(64 a^4+m_1^2 m_2^2 (1-2 \sigma ^2\right)^2)}{128 \pi  m_1 m_2 \left(\sigma ^2-1\right)^{3/2} m_p^4} \log{(\boldsymbol{q}^2_\perp )}  \\
        & + \frac{\left(m_1+m_2\right) \left(128 a^4-4 a^2 m_1 m_2 \left(\sigma ^2-17\right)+3 m_1^2 m_2^2 \left(5 \sigma ^2-1\right)\right)-512 a^2 b m_1 m_2}{512 m_p^4 |\boldsymbol{q}_\perp|} \\
        & -\frac{e^2 m_1 m_2 \texttt{g}_c \left(4 \sqrt{2} a \left(\alpha _1^2+\alpha _2^2\right)+4 \alpha _1 \alpha _2 \sigma  \left(3 \left(m_1+m_2\right)-4 \sqrt{2} a\right)+\left(3 \sigma ^2-1\right) \left(\alpha _1^2 m_2+\alpha _2^2 m_1\right)\right)}{16 m_p^2 |\boldsymbol{q}_\perp|} \\
        &  +\frac{4 \alpha _1^2 \alpha _2^2 e^4 \left(m_1+m_2\right) \texttt{g}_c^2}{|\boldsymbol{q_\perp}|}\,.
        \label{del1p}
    \end{split}
\end{align}
Note that the $\mathcal{O}(|\boldsymbol{q}_\perp|^{-2}) $ term of the 1-loop amplitude is completely canceled at $\epsilon^0$ order by the tree convolution. We also note that the $\delta^{(1)}$ has an imaginary part. However, the imaginary part of the eikonal phase does not contribute to the conservative scattering angle. Physically, the imaginary part of the eikonal phase measures the number of massless particles emitted from the two-body system. We now proceed to the computation of the eikonal phase in impact-parameter space by taking the Fourier transform. 
\begin{align}
    \delta(\sigma, \boldsymbol{b}_e )=\frac{1}{N} \int\frac{d^{D-2}\boldsymbol{q}_\perp}{(2 \pi)^{D-2}} e^{i \boldsymbol{b}_e \cdot \boldsymbol{q}_\perp} \delta(\sigma,\boldsymbol{q}_\perp)\,.
\end{align}
So using \eqref{del0p} and \eqref{del1p} respectively we have
\begin{flalign}
   \hspace{-0 cm}  \delta^{(0)}(\sigma, \boldsymbol{b}_e )&= \left( -\frac{a^2}{4 \pi  \sqrt{\sigma ^2-1} m_p^2}+\frac{\alpha _1 \alpha _2 e^2 \sigma  \texttt{g}_c}{\pi  \sqrt{\sigma ^2-1}}-\frac{m_1 m_2 \left(2 \sigma ^2-1\right)}{32 \pi  \sqrt{\sigma ^2-1} m_p^2} \right) \log(\boldsymbol{b}^2_e)&
\end{flalign}
and,
\begin{align}
    \begin{split}
        \delta^{(1)}(\sigma, \boldsymbol{b}_e ) &= \frac{128 a^4 \left(m_1+m_2\right)-4 a^2 m_1 m_2 \left(128 b+\left(m_1+m_2\right) \left(\sigma ^2-17\right)\right)+3 m_1^2 m_2^2 \left(m_1+m_2\right) \left(5 \sigma ^2-1\right)}{4096 \pi  m_1 m_2 \sqrt{\sigma ^2-1}  m_p^4 |\boldsymbol{b}_e| } \\ & -\frac{e^2 \texttt{g}_c \left(4 \sqrt{2} a \left(\alpha _1^2+\alpha _2^2\right)+4 \alpha _1 \alpha _2 \sigma  \left(-4 \sqrt{2} a+3 m_1+3 m_2\right)+\left(3 \sigma ^2-1\right) \left(\alpha _1^2 m_2+\alpha _2^2 m_1\right)\right)}{128 \pi  \sqrt{\sigma ^2-1} m_p^2 |\boldsymbol{b}_e|} \\
        & + \frac{\alpha _1^2 \alpha _2^2 e^4 \left(m_1+m_2\right) \texttt{g}_c^2}{2 \pi  m_1 m_2 \sqrt{\sigma ^2-1} |\boldsymbol{b}_e|}\,.
    \end{split}
\end{align}
\textbf{\textit{Scattering angle:}}\\
The phase in impact parameter space and the amplitude in momentum space are related as
\begin{align}
    \mathbfcal{A}(\sigma,-q^2)=\int d^{D-2} \boldsymbol{b}_e \left( e^{i \delta(\sigma,\boldsymbol{b}_e)} -1 \right) e^{-i \boldsymbol{b}_e \cdot \boldsymbol{q}} 
\end{align}
and stationary phase approximation yields the relation
\begin{align}
    \boldsymbol{q}=-\frac{\partial}{\partial\boldsymbol{b}_e} \delta(\sigma,\boldsymbol{b}_e)\,. \label{5.14}
\end{align}
In the centre-of-mass frame, $\boldsymbol{q}$, scattering angle $\chi$ and the COM momentum $\boldsymbol{p}$ are related as 
\begin{align}
    | \boldsymbol{q}|=2 |\boldsymbol{k}| \sin{\frac{\chi}{2}} \label{5.15}
\end{align}
where in terms of centre-of-mass energy
\begin{align}
    |\boldsymbol{k}|=\frac{m_1 m_2 \sqrt{\sigma^2 - 1}}{E} = \frac{m_1 m_2 \sqrt{\sigma^2 - 1}}{\sqrt{m_1^2 + m_2^2 +2 m_1 m_2 \sigma}}\,.
\end{align}
So from \eqref{5.14} and \eqref{5.15} we have
\begin{align}
    \sin{\frac{\chi}{2}}=-\frac{1}{2 |\boldsymbol{k}|} \frac{\partial}{\partial |\boldsymbol{b}_e|} \Re \delta(\sigma,\boldsymbol{b}_e)\,. \label{5.17}
\end{align}
Now we can write RHS of \eqref{5.17} by order
\begin{flalign}
    \frac{\chi^{(0)}}{2}&=\frac{a^2}{4 \pi  |\boldsymbol{k}||\boldsymbol{b}_e| \sqrt{\sigma ^2-1} m_p^2} - \frac{\alpha _1 \alpha _2 e^2 \sigma  \texttt{g}_c}{\pi  |\boldsymbol{k}||\boldsymbol{b}_e| \sqrt{\sigma ^2-1}}+\frac{m_1 m_2 \left(2 \sigma ^2-1\right)}{32 \pi  |\boldsymbol{k}||\boldsymbol{b}_e| \sqrt{\sigma ^2-1} m_p^2}\,, &
\end{flalign}

\begin{align}
    \begin{split}
        \frac{\chi^{(1)}}{2} &= \frac{a^4 \left(m_1+m_2\right)}{64 \pi  m_1 m_2 |\boldsymbol{k}||\boldsymbol{b}^2_e| \sqrt{\sigma ^2-1} m_p^4}-\frac{a^2 \left(128 b+\left(m_1+m_2\right) \left(\sigma ^2-17\right)\right)}{2048 \pi  |\boldsymbol{k}||\boldsymbol{b}^2_e| \sqrt{\sigma ^2-1} m_p^4}  +\frac{3 m_1 m_2 \left(m_1+m_2\right) \left(5 \sigma ^2-1\right)}{8192 \pi  |\boldsymbol{k}||\boldsymbol{b}^2_e| \sqrt{\sigma ^2-1} m_p^4} \\
        & -\frac{e^2 \texttt{g}_c \left(4 \sqrt{2} a \left(\alpha _1^2+\alpha _2^2\right)+4 \alpha _1 \alpha _2 \sigma  \left(-4 \sqrt{2} a+3 m_1+3 m_2\right)+\left(3 \sigma ^2-1\right) \left(\alpha _1^2 m_2+\alpha _2^2 m_1\right)\right)}{256 \pi  |\boldsymbol{k}||\boldsymbol{b}^2_e| \sqrt{\sigma ^2-1} m_p^2} \\
        & + \frac{\alpha _1^2 \alpha _2^2 e^4 \left(m_1+m_2\right) \texttt{g}_c^2}{4 \pi  m_1 m_2 |\boldsymbol{k}||\boldsymbol{b}^2_e| \sqrt{\sigma ^2-1}}\,.
    \end{split}
\end{align}
The above results can be written in terms of angular momentum 
\begin{align}
    J=|\boldsymbol{b} \times \boldsymbol{k}|,
\end{align}
where $\boldsymbol{b}$ is the impact parameter perpendicular to the incoming centre-of-mass momentum $\boldsymbol{p}$. Note, however, this impact parameter is different from the $\boldsymbol{b}_e$ occurring in the eikonal phase, which points in the direction of momentum transfer $\boldsymbol{q}$. The magnitude of $\boldsymbol{b}_e$ and $\boldsymbol{b}$ are related by
\begin{align}
    |\boldsymbol{b}|=|\boldsymbol{b}_e| \cos{\frac{\chi}{2}}.
\end{align}
This difference is unimportant for small-angle scattering, and it will only matter at order $J^{-3}$. So in terms of $J$ and $G$ ($ m_p=(32 \pi G)^{-1/2}$)  we have,
\begin{flalign}
    \frac{\chi^{(0)}}{2}&=\frac{8 a^2 G}{J \sqrt{\sigma ^2-1}}-\frac{\alpha _1 \alpha _2 e^2 \sigma  \texttt{g}_c}{\pi  J \sqrt{\sigma ^2-1}}+\frac{G m_1 m_2 \left(2 \sigma ^2-1\right)}{J \sqrt{\sigma ^2-1}}\,, &
\end{flalign}

\begin{align}
    \begin{split}
        \frac{\chi^{(1)}}{2}&= \frac{16 \pi  a^4 G^2 \left(m_1+m_2\right)}{J^2 \sqrt{2 m_2 m_1 \sigma +m_1^2+m_2^2}} -\frac{\pi  a^2 G^2 m_1 m_2 \left(128 b+\left(m_1+m_2\right) \left(\sigma ^2-17\right)\right)}{2 J^2 \sqrt{2 m_2 m_1 \sigma +m_1^2+m_2^2}} \\
        & + \frac{3 \pi  G^2 m_1^2 m_2^2 \left(m_1+m_2\right) \left(5 \sigma ^2-1\right)}{8 J^2 \sqrt{2 m_2 m_1 \sigma +m_1^2+m_2^2}} +\frac{\alpha _1^2 \alpha _2^2 e^4 \left(m_1+m_2\right) \texttt{g}_c^2}{4 \pi  J^2 \sqrt{2 m_2 m_1 \sigma +m_1^2+m_2^2}} \\
        & -\frac{e^2 G m_1 m_2 \texttt{g}_c \left(4 \sqrt{2} a \left(\alpha _1^2+\alpha _2^2\right)+4 \alpha _1 \alpha _2 \sigma  \left(-4 \sqrt{2} a+3 m_1+3 m_2\right)+\left(3 \sigma ^2-1\right) \left(\alpha _1^2 m_2+\alpha _2^2 m_1\right)\right)}{8 J^2 \sqrt{2 m_2 m_1 \sigma +m_1^2+m_2^2}}\,.
    \end{split}
\end{align}
If we `turn off' the coupling, i.e $G \rightarrow 0$ \cite{Bern:2021xze} or $e \rightarrow0$ \cite{Mogull:2020sak}, and set the screening constants $a_i \, ,  b_i =0$, we retrieve the known gravitational and (scalar) electromagnetic scattering angles. 
\section{Conclusions and Discussion}
\label{sec:conclusions}
We have analyzed classical scattering of charged, non-spinning compact objects in EMD theory using modern scattering–amplitude methods, with cross–checks based on potential theory. Our goal was to extract the conservative two–body potential, establish the infrared (IR) structure and its cancellation in momentum space, and compute the eikonal phase and associated scattering angle through one loop (classical $2{\rm PM}$ order).

\begin{itemize}
  \item  We have motivated EMD as a low–energy effective description inspired by heterotic string theory, in which additional long–range fields (dilaton and gauge bosons) arise from first principles. This provides a UV-motivated setting in which higher–curvature corrections and extra degrees of freedom naturally coexist, while our concrete computations focused on the leading two-derivative EMD sector.
  \item After fixing conventions and gauges, we derived the propagators and Feynman rules needed for $2\!\to\!2$ scattering. This delivers a set of building blocks separating purely gravitational, electromagnetic, and dilatonic exchanges and their interference.
  \item  Using dimensional regularization, expansion by regions, and IBP reduction, we obtained a basis of master integrals and their soft (classical) expansions that capture the long–range, nonanalytic in momentum transfer relevant for conservative dynamics. These ingredients were assembled into the one–loop amplitudes across the relevant topologies (single–exchange, triangle, box, and penguin).
  \item We computed the momentum–space two–body potential via the Lippmann-Schwinger equation. \textit{A careful EFT/Born subtraction removes iterated long–range exchanges, and we explicitly showed that the resulting momentum–space potential is IR finite, mirroring the situation in GR}. This establishes that the soft amplitude, once dressed by the appropriate iterative counter term, is free of IR divergences at the level relevant for conservative observables.
  \item  The conservative scattering angle was obtained by exponentiating the momentum–space amplitude (eikonal exponentiation) and differentiating the eikonal phase with respect to impact parameter, and the result is given as an explicit closed-form function of gravitational, electromagnetic, and dilatonic couplings. We also verify that, in the appropriate limits, our results for the classical potential and the scattering angle agree precisely with those available in the literature.

\end{itemize}
\textbf{Future outlook}:
There are several interesting directions to be done. 
\begin{itemize}
  \item \emph{Higher orders.} \textcolor{black}{Extending to two loops ($3{\rm PM}$) would sharpen the conservative sector and test the robustness of IR cancellations beyond one loop in EMD. However, at two-loop and onward radiative corrections appear to emerge. However, we think that, as long as we consider the conservative scattering angle at two loops, the Born subtraction scheme should work. However if we add the radiative corrections (where one needs to compute the boundary integrals both from potential and radiation region) the subtraction scheme may fail. However, in those cases, the EFT prescription has been checked to work perfectly in the context of GR in \cite{Cheung:2018wkq}. We hope to report this soon \cite{arpan2}.}
  \item \emph{Spin and finite size.} Incorporating spins, spin–orbit/spin–spin couplings, and tidal responses (including dilatonic/electromagnetic polarizabilities) is essential for realistic compact objects and will enrich the phenomenology.
  \item \emph{Radiation and reaction.} Computing real emission (gravitational, electromagnetic, and dilatonic) and the accompanying radiation–reaction forces (in line with \cite{Caron-Huot:2023vxl}) will complete the bridge to waveform modeling for generic scattering and bound–state transitions, which is also work in progress \cite{arpan1}.
  \item \emph{Higher–curvature corrections.} Including $\alpha'$ \cite{Metsaev:1987zx}(higher–derivative) operators from the stringy effective action will test how UV-motivated terms imprint themselves on classical observables and whether symmetry protection persists at higher PM orders.
\end{itemize}

\section*{Acknowledgments}
AB is supported by the Core Research Grant (CRG/2023/005112) by DST-ANRF of the India Govt.  AB also acknowledges the associateship program of the Indian Academy of Science, Bengaluru. SG (PMRF ID: 1702711) and SP (PMRF ID: 1703278) are supported by the Prime Minister’s Research Fellowship of the Government of India. AM is supported by the Sabarmati Bridge Fellowship (OTH/MoE/13474) provided by the Indian Institute of Technology Gandhinagar.
\bibliography{ref}
\bibliographystyle{utphysmodb}
\end{document}